\documentclass[aps,pra,twocolumn,showpacs,superscriptaddress,10pt,longbibliography]{revtex4-2}
\usepackage{float}
\usepackage{multirow}
\usepackage{amsmath}
\usepackage{amssymb}
\usepackage{graphicx}% Include figure files
\usepackage{graphics}
\usepackage{dcolumn}% Align table columns on decimal point
\usepackage{bm}% bold math
\usepackage{hyperref}
\usepackage{xcolor}

\usepackage{mathptmx}
\usepackage{etoolbox}
\newcommand{\scs}{\scriptscriptstyle}
\newcommand{\smallm}{\scs (\!-\!)}
\newcommand{\smallp}{\scs (\!+\!)}
\newcommand{\smallmp}{\scs (\!\mp\!)}
\newcommand{\smallpm}{\scs (\!\pm\!)}
\newcommand{\sml}{\scs{\leftarrow}}
\newcommand{\smr}{\scs{\rightarrow}}
\graphicspath{{Fig/}} % Figures pathway

\begin{document}

% \title{Bound states in the continuum in 2D metasurfaces?}
\title{Temporal coupled mode theory: from bound states in the continuum to uniguided resonances}

\author{Dmitrii N. Maksimov}
\affiliation{Qingdao Innovation and Development Center, Harbin Engineering University, Qingdao 266000, Shandong, China}
\affiliation{Kirensky Institute of Physics, Federal Research Centre KSC SB RAS, 660036, Krasnoyarsk, Russia}
\affiliation{IRC SQC, Siberian Federal University, 660041, Krasnoyarsk, Russia}
%\email{mdn@tnp.krasn.ru}

\author{Pavel S. Pankin}
\email{pavel-s-pankin@iph.krasn.ru}
\affiliation{Qingdao Innovation and Development Center, Harbin Engineering University, Qingdao 266000, Shandong, China}
\affiliation{Kirensky Institute of Physics, Federal Research Centre KSC SB RAS, 660036, Krasnoyarsk, Russia}
\affiliation{School of Engineering Physics and Radio Electronics, Siberian Federal University, 660041, Krasnoyarsk, Russia}

\author{Dong-Wook Kim}
\affiliation{Qingdao Innovation and Development Center, Harbin Engineering University, Qingdao 266000, Shandong, China}
\affiliation{School of Physics and Engineering, ITMO University, St. Petersburg 197101, Russia}

\author{Mingzhao Song}
\affiliation{Qingdao Innovation and Development Center, Harbin Engineering University, Qingdao 266000, Shandong, China}

\author{Chao Peng}
\affiliation{State Key Laboratory of
Photonics and Communications, School of Electronics, \& Frontiers Science Center for Nano-optoelectronics, Peking University, Beijing 100871, China}

\author{Andrey A. Bogdanov}
\email{a.bogdanov@hrbeu.edu.cn}
\affiliation{Qingdao Innovation and Development Center, Harbin Engineering University, Qingdao 266000, Shandong, China}
\affiliation{School of Physics and Engineering, ITMO University, St. Petersburg 197101, Russia}
\email{a.bogdanov@hrbeu.edu.cn}

\date{\today}

\begin{abstract}
We revise the temporal coupled mode theory to describe the resonant response of quasi-guided modes under
oblique incidence in periodic dielectric metasurfaces accounting for the constraints imposed by the symmetry of
the unit cell. We derive a universal expression for the Fano resonance line shape in reflection and transmission
for subwavelength period metasurfaces and reveal a fundamental link between the Fano parameters and unit cell
symmetries, including robust reflection and transmission zeros, as well as spectral signatures of bound states
in the continuum. To demonstrate the power of the developed theory, we provide a generalized framework
for analyzing asymmetry in the coupling and decoupling coefficients. This allows us to reveal the dual nature
of unidirectional guided resonances and counter-propagating modes characterized by single-sided coupling to
incident waves, which have been actively discussed in recent years. These insights provide a powerful analytical
tool for designing high-Q resonances in metasurfaces through symmetry exploitation.
\end{abstract}

\maketitle
%%%%%%%%%%--INTRODUCTION---%%%%%%%%%%%%%%%%%%%
\newpage
%\tableofcontents

\newpage
\section{Introduction} \label{Intro}

In recent years dielectric metasurfaces have become an indispensable tool in the fields of photonics and radiophysics since their application paves the way to shaping optical response induced by high-quality (high-$Q$) modes \cite{Kuznetsov2016, Campione2016, genevet2017recent, Algorri2019, Koshelev2020}. The propagating modes in periodic dielectric metasurfaces are Bloch waves traveling across the plane of the metasurface. If the frequency of such waves at a given in-plane Bloch wavenumber occurs above the line of light, the modes radiate electromagnetic energy into outer space. Thus, the modes become leaky and, as such, can be termed {\it quasi-guided} or {\it radiative} modes \cite{tikhodeev2002quasiguided, Huang2023}. On the other hand, leaky modes are {\it resonant}, that is, can be coupled with incident light to induce a resonant response, which is accompanied by the enhancement of the electromagnetic field in the host structure \cite{Mocella15, yuan2017strong}. This phenomenon has been instrumental in the design of various devices that employ a strong light-matter interaction, even with low-loss dielectrics \cite{baranov2017all, Kivshar2018}. Nowadays, such devices include lasers \cite{Kodigala17, shankhwar2017high, Huang2020, yang2021low, azzam2021single,  Hwang2022}, sensors \cite{Romano18b, Ollanik2018, Yesilkoy19, Tseng2020, Wang2021, Conteduca2021, Liu2023}, absorbers \cite{Liu2017, Fan2017, Saadabad21, Tian2018, Chen2019}, polaritonic devices~\cite{Al-Ani2022Jan,Wu2024Apr,Sortino2024Jul,Masharin2024Jan}, and nonlinear optical components~\cite{Yang2015,Schiattarella2024Feb,Liu2019Dec,Kravtsov2020Apr,Koshelev2019Jul}.
Formally, various types of resonances in metasurfaces, including Wood anomalies~\cite{Sarrazin2003Feb,Wood1935Dec}, lattice resonances~\cite{Kravets2018Jun,Castellanos2019Jun}, quasi-bound states in the continuum~\cite{Rybin17a}, and high-$Q$ Mie resonances~\cite{Babicheva2024Sep} can be associated with quasi-guided resonant modes also know as {\it resonant states}~\cite{Doost2014Jul,Both2021Dec} or {\it quasi-normal modes}~\cite{Lalanne2018May,Alpeggiani2017Jun}.

In the far field, the high-$Q$ resonant modes manifest themselves as sharp Fano resonances in the transmittance spectrum \cite{Campione2016, zhou2014progress, Limonov17, Krasnok19}. The Fano resonances are typically understood as a product of interference between waves following two optical paths. The above interpretation can be applied to dielectric metasurfaces via decomposing the electromagnetic field into a resonant field, which is induced by coupling to a resonant eigenmode, and a background field, which is almost independent of the incident frequency on the scale of the high-$Q$ resonance width \cite{Fan03}. The Fano resonances, therefore, are the primary signature of the resonant modes in a dielectric metasurface. The intricate behavior of Fano resonances makes it possible to apply dielectric metasurfaces to shape the transmittance spectrum for filtering \cite{Foley15, monti2020design}, polarization conversion \cite{Guo17, hu2020all}, and circular dichroism \cite{gorkunov2020metasurfaces,toftul2024chiral}.

One remarkable feature of the subwavelength dielectric metasurfaces is the opportunity to control the life-time of the resonant modes. In the ideal case of a lossless dielectric structures, non-radiative modes embedded within the continuum of scattering states can attain infinite lifetimes even when their frequencies lie above the light line. Such modes are known as bound states in the continuum (BICs) and are representative of the utmost case of light localization \cite{Hsu16, koshelev2019meta, joseph2021bound, Kang2023}. By definition, the BICs are not coupled with incident radiation, and, therefore, are seemingly useless in applications. However, BICs can be activated by breaking the system's symmetry, either structurally \cite{Koshelev18}, or by incident field \cite{Maksimov20}. As a consequence, the BIC can be transformed into a quasi-BIC with extremely high quality factor ($Q$-factor) which diverges to infinity as the symmetry is recovered under variation of a certain control parameter, e.g., the angle of incidence. The $Q$-factor divergence manifests itself in the transmittance spectrum as a collapsing Fano resonance with  the line width vanishing on approach to the BIC point in the parametric space \cite{Blanchard16, bogdanov2019bound, pankin2020fano, Bulgakov18b}. The collapse of the Fano resonance is complemented by the critical field enhancement in the lossless case \cite{Mocella15, Yoon2015Dec}. In the presence of material absorption, precise control over the radiative Q factor allows to achieve the critical coupling regime between q-BICs and the incident wave characterized by the maximal field enhancement~\cite{Yoon2015Dec}.

Another interesting type of high-$Q$ resonant modes that have recently attracted a lot of attention in the literature is a unidirectional guided resonant mode (UGR) \cite{yin2020observation}. This type of resonant mode can be viewed as a semi-BIC since the UGRs do not radiate only to a single direction, i.e., either to the substrate or to the superstrate of the metasurface. In addition to intriguing topological properties \cite{Jiang2023, Yin2023, Zhuang2024, Wang2024}, UGRs are applicable to optical devices that utilize asymmetric coupling between the resonant mode and the outgoing waves \cite{Zhang2021, Zhang2022, Li2023, Xu2023, Xu2023a, Wang2024a,hidalgo2025cloaked}.   

Theoretical description of the Fano resonances induced by various types of quasi-guided modes in dielectric metasurfaces is hindered by the complexity of the host structures. The aforementioned phenomena occur only in dielectric systems, which are at least two-dimensional (2D) structures with the permittivity modulated in one spatial dimension. This does not allow for analytic solutions of Maxwell's equations
to be written in a closed form. Nonetheless, it is possible to derive a generic expression for the transmittance as a function of the incident frequency by using temporal coupled mode theory (TCMT). The TCMT is a phenomenological instrument 
that relies on symmetries and conservation laws to establish constraints on the number of parameters that describe the Fano line-shape \cite{Fan03, Suh04}. %\textcolor{red}
The TCMT approach has proved to be efficient for analyzing the scattering and absorption spectra of dielectric metasurfaces \cite{Zhou2016, Alpeggiani17, Ming2017, maksimov2020optical, Bikbaev21, Zhang2023, Wu2022, Huang2024}. However, the TCMT in its conventional form treats the coupling and decoupling parameters as fitting constants, and therefore ignores the rigorous constraints imposed by unit-cell symmetry. This limitation becomes pivotal when the resonant state is decoupled from one or more scattering channels, such as BICs or UGR.

In this work, we revisit the TCMT in application to the Fano response from dielectric metasurfaces hosting high-$Q$ resonant modes. The developed approach rigorously describes various types of high-$Q$ resonances, including BICs and UGRs, and reflects the role of the symmetry of the metasurface unit cell. We derive a generic formula for the line-shape of the Fano resonance in transmittance for the lossless metasurfaces. We analyze metasurfaces with various unit cell symmetries and uncover the effects that symmetry incurs on the profile of the Fano resonance induced by an isolated high-$Q$ mode.  

The article is organized as follows. In Sec.~\ref{Sec2} we introduce the system under scrutiny and discuss the generic properties of the scattering matrix ($S$-matrix) in the regime of specular reflection. In Sec.~\ref{Sec3} we overview and revisit the TCMT to derive the generic form of the constraints imposed on the resonant mode coupling and decoupling coefficients by energy conservation and time-reversal symmetry. In Sec.~\ref{Sec4} we apply the results of the previous section to dielectric metasurfaces and derive the generic TCMT solution for the Fano resonance line-shape. In Sec.~\ref{Sec5} we verify our findings numerically in comparison with finite-difference frequency domain (FDFD)~\cite{rumpf2022electromagnetic} simulations for all types of point group symmetries of the elementary cell including in-plane mirror symmetry, out-of-plane mirror symmetry, and inversion symmetry. In Sec.~\ref{Sec6} we apply our approach to UGRs. It is shown for the asymmetric metasurfaces that the UGR is dual to a counter-propagating mode of a peculiar type. Unlike the UGR which is not coupled with one of the outgoing channels whereas being coupled with both incident ones, the dual mode, which we term unidirectional coupled resonant mode (UCR), is not coupled with one of the incident channels whereas is still coupled with both outgoing ones. Finally, we summarize and conclude in Sec.~\ref{Sec7}.

%%%%%%%%%%%%%%%%%%%%%%%%%%%%%%%%%%%%%%
%%%%--FIG1---%%%%%%%%%%%%%%%%%
%%%%%%%%%%%%%%%%%%%%%%%%%%%%%%%%%%%%%%%%%%%%%%%

\begin{figure}[t]
    \centering
    \includegraphics[width=8cm]{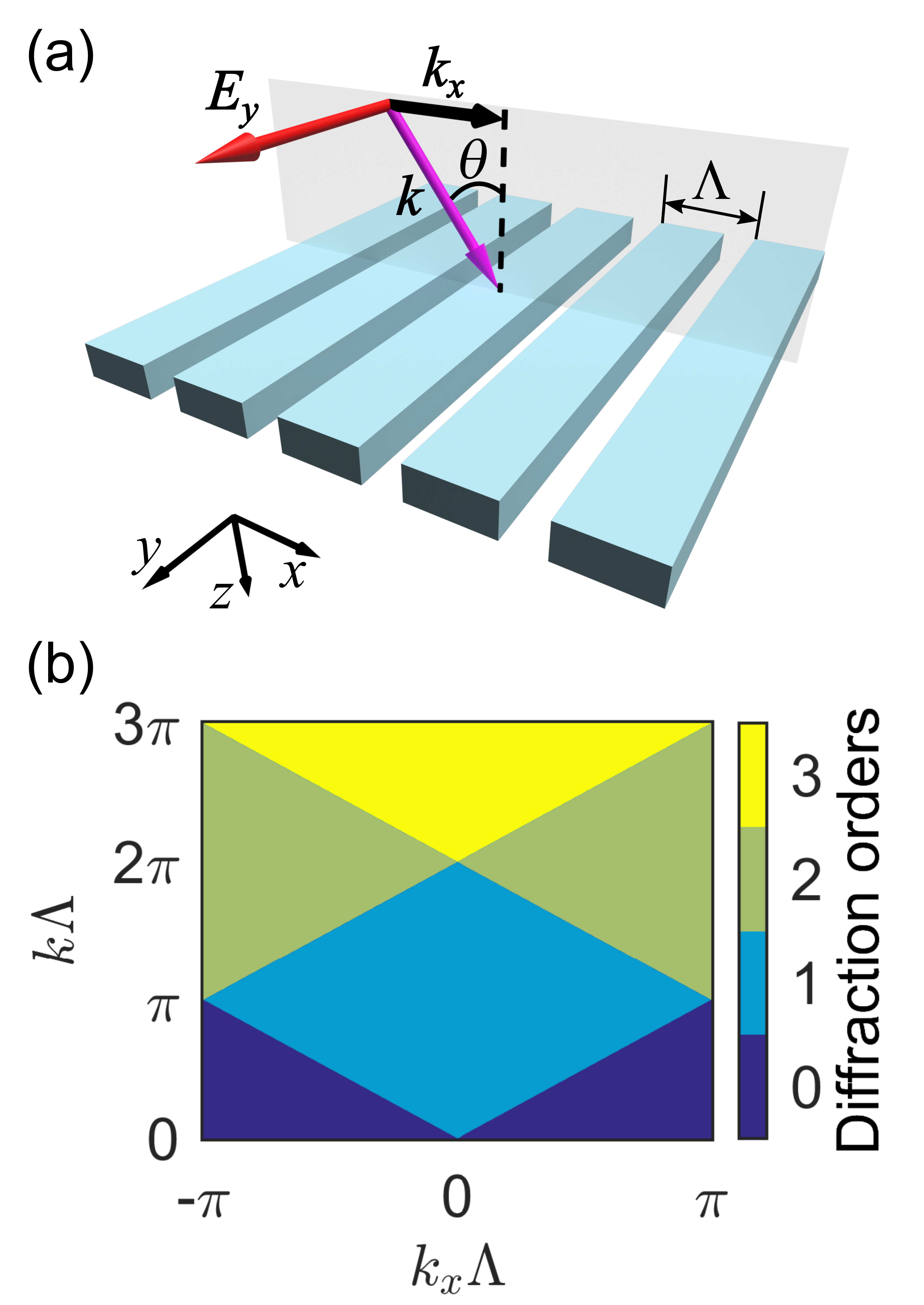}

\caption{Dielectric metasurface in the form of a ruled grating. (a) The light blue bars show the ruled grating with refractive index $n_1 = 3.5$. The plane of incidence is shaded in gray. The magenta arrows shows the wave vector $k$ of the incident field. The $y$-component of the electric field and the $x$-component of $k$ are shown as vectors. The refractive index of ambient medium $n_0 = 1.5$. (b) Diagram of diffraction orders. The colorbar on the right indicates the number of channels open for diffraction on either side of the metasurface.}
    \label{Fig1}
\end{figure}

\section{System description}\label{Sec2}

The system to be considered is shown in Fig.~\ref{Fig1}. One can see in Fig.~\ref{Fig1}~(a) that the system is a ruled grating that possesses a translation symmetry in the $x$-direction with period $\Lambda$. The body of the grating is made of lossless dielectric with permittivity $n_1=3.5$. As it has been already mentioned before, here we focus on 2D electrodynamics which means that all fields are independent of $y$. This can be achieved by choosing the $x0z$-plane as the plane of incidence. Further on we focus on the transverse electric wave (s-waves) with the electric vector perpendicular to the plane of incidence. Albeit the scattering of light is controlled by vectorial Maxwell's equations the problem can be reduced to a single scalar equation for the $y$-component of the electric field while the other components are equal to zero
\begin{equation}\label{Helholtz}
\left(\frac{\partial^2}{\partial x^2}+\frac{\partial^2}{\partial z^2}\right)E_y(x,z)+\epsilon(x,z)\frac{\omega^2}{c^2}E_y(x,z)=0,
\end{equation}
where $\epsilon(x,z)$ is the dielectric function, $\omega$ is the angular frequency of the incident light, and $c$ is the speed of light. In the far-field the dielectric function is constant, thus, the monochromatic solutions are plane waves
\begin{equation}\label{11}
E_y(x,z)\propto e^{\pm ik_x x\pm ik_z z}, \ \ \frac{n_0^2\omega^2}{c^2}=k_x^2+k_z^2,
\end{equation}
where $n_0=1.5$ is the refractive index of the ambient medium, and $k_{x, z}$ are the corresponding components of the wave vector. Since the grating is periodic the solution has to comply with the Bloch theorem
\begin{align}\label{12}
 E_y(x,z)=E_{y}^{\scs(0)}(x,z)e^{i\beta x}, 
%& E^{\scs(0)}_{y}(x,z)=E_{y}^{\scs(0)}(x+m\Lambda,z) %\nonumber \\
%& m=-\infty, \ldots, -1, 0, 1, \ldots, \infty, \nonumber \\
\ \ \beta\Lambda\in(-\pi, \pi],
\end{align}
where $E^{\scs(0)}_{y}(x,z)$ is a periodic function of $x$ with period $\Lambda$, and $\beta$ is the Bloch propagation constant. The conservation of quasi-momentum requires
$k_x=\beta+{2\pi m}/{\Lambda}$ where index $m$ is known as the diffraction order. By setting $k_z=0$ one finds the cut-off frequencies for the diffraction scattering channels as follows
\begin{equation}
n_0\omega=\pm c\left(\beta+\frac{2\pi m}{\Lambda}\right), \ \ m=-\infty, \ldots, -1, 0, 1, \ldots, \infty.
\end{equation}
The diagram of the diffraction channels is shown in Fig.~\ref{Fig1}~(b).
The specular reflection occurs when only the zeroth-order diffraction channel is open for propagation of light, i.e.
\begin{equation}\label{spec}
|k_x|<\frac{n_0\omega}{c}<\frac{2\pi}{\Lambda}-|k_x|.
\end{equation}

Because the modes above the light are radiative they are described by complex eigenfrequencies $ \omega_{0}-i\gamma $, where $\gamma$ is the radiative decay rate. Due to reciprocity, every propagating solution is doubly degenerate: a forward travel mode with Bloch wavenumber $\beta$ always has a counter-propagating partner with wavenumber $-\beta$ with the same complex eigenfrequency. We denote their amplitudes by $a_{\smr}$ and $a_{\sml}$, respectively, with the arrows indicating the direction of propagation.

\section{S-matrix for oblique incidence}
\begin{figure}[t]
    \centering    \includegraphics[width=1\linewidth]{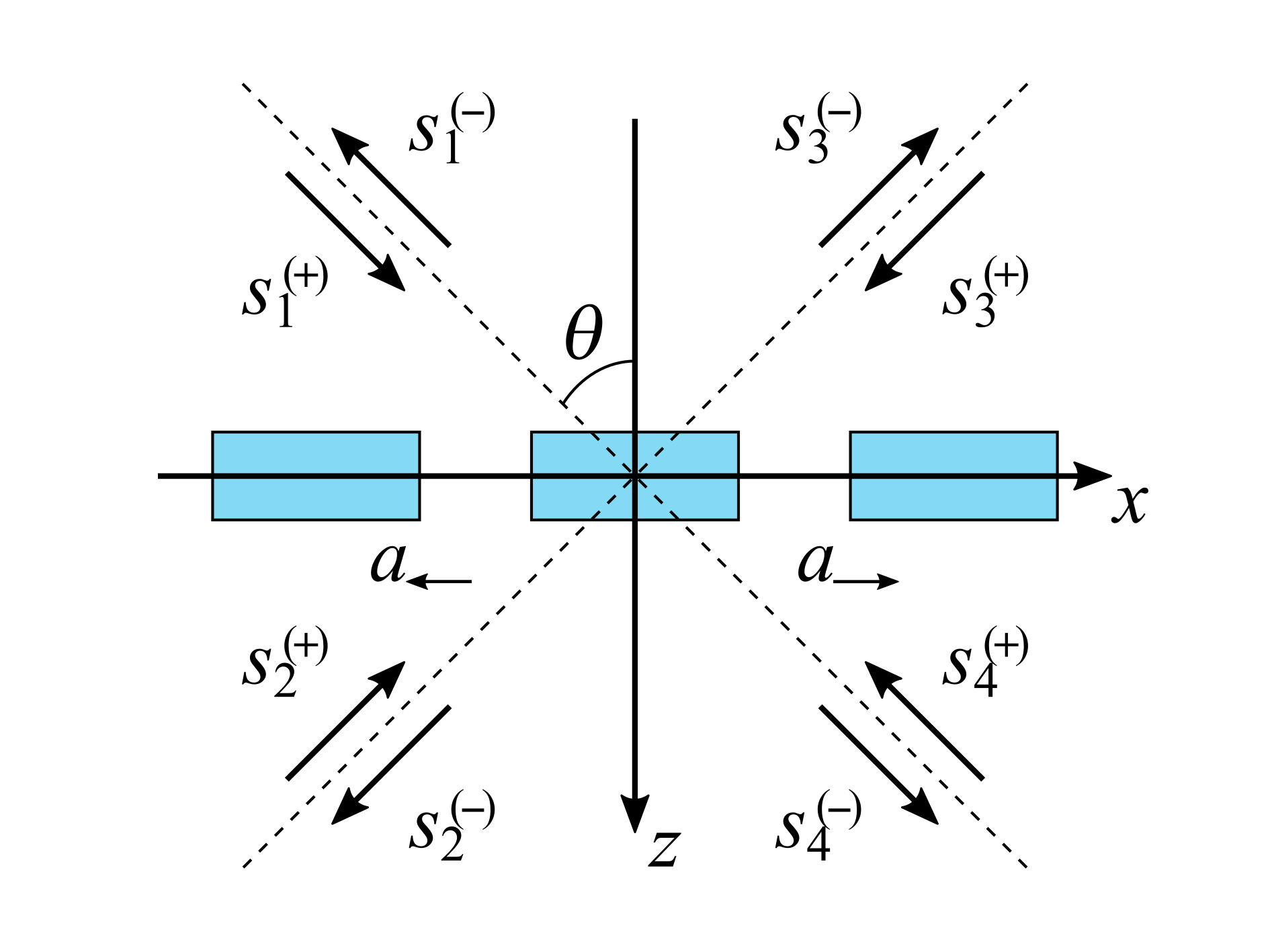}    %\includegraphics[width=1\linewidth]
    %{Grating_Types.jpg} 
    \caption{Definition of the scattering channels for the structure shown in Fig.~\ref{Fig1}~(a). $\theta$ is the angle of incidence.}\label{Fig2}   
\end{figure}

We consider scattering of plane monochromatic waves in the spectral domain where only specular reflection is possible, see Eq.~\eqref{spec}.
This assumption, together with the specified incident frequency $\omega$ and the angle of incidence $\theta$, drastically simplify our analysis, since only four scattering channels are open
for incident/outgoing light. The scattering channels are defined in Fig.~\ref{Fig2}, where the superscript $^{\smallp}$ is used for the incident channels and $^{\smallm}$ for the outgoing ones.
We also assume that the outgoing channels are linked to the incident ones via the time-reversal operation. 
The $4\!\times\!4$ $S$-matrix is implicitly defined through the following formula that relates the vectors of incident and outgoing (scattered) amplitudes
\begin{equation}\label{Smatrix}
{\bf s}_{\smallm}
=\widehat{S}_{\scs{4\!\times\!4}}
{\bf s}_{\smallp},
\end{equation}
where the vectors of outgoing ${\bf s}^{\smallm}$ and incident ${\bf s}^{\smallp}$ amplitudes are given by
\begin{equation}
\bf{s}_{\scs{(\pm)}}=
\left(
\begin{array}{c}
s_1^{\scs{(\pm)}} \\
s_2^{\scs{(\pm)}} \\
s_3^{\scs{(\pm)}} \\
s_4^{\scs{(\pm)}}
\end{array}
\right).
\end{equation}
%${\bf{s}^{\scs{(-)}}}^{*}=\hat{\mathcal T}\bf{s}^{\scs{(+)}}$.

Note that in case $\theta\neq0$ the time-reversal changes the propagation direction of all waves with respect to the $x$-axis, i.e. $\beta\rightarrow-\beta$. If follows from the energy conservation that ${{\bf s}_{\smallm}}^{\dagger}{\bf s}_{\smallm}={\bf s}_{\smallp}^{\dagger}{\bf s}_{\smallp}$ and the scattering matrix is unitary 
\begin{equation}\label{0A2}
\widehat{S}_{\scs{4\!\times\!4}}^{-1}=\widehat{S}_{\scs{4\!\times\!4}}^{\dagger}.
\end{equation}
For time-reversal systems,  ${\bf s}_{\smallpm}={\bf s}_{\smallmp}^{*}$ and the scattering matrix satisfies 
\begin{equation}\label{0A21}
\widehat{S}_{\scs{4\!\times\!4}}^{-1}=\widehat{S}_{\scs{4\!\times\!4}}^{*}.
\end{equation}
Time-reversal invariance, combined with energy conservation, enforces reciprocity and thus renders the scattering matrix symmetric.    
\begin{equation}\label{0A1}
\widehat{S}_{\scs{4\!\times\!4}}=\widehat{S}_{\scs{4\!\times\!4}}^{\intercal}.
\end{equation}

In addition, the $S$-matrix must comply with the momentum conservation which together with unitarity and symmetry of the $S$-matrix yields
\begin{equation}\label{0A3}
\widehat{S}_{\scs{4\!\times\!4}}=
\left(
\begin{array}{cc}
0 & \widehat{S} \\
\widehat{S}^{\intercal} & 0
\end{array}
\right),
\end{equation} 
where $\widehat{S}$ is a $2\!\times\!2$ unitary matrix.
Note that the matrix $\widehat{S}$ does not have to be symmetric. At this point we can only apply the generic expression for the $2\!\times\!2$ unitary matrix
\begin{equation}\label{S2gen}
\widehat{S}=e^{i\psi}
\left(
\begin{array}{cc}
\rho e^{i\eta} & i\tau e^{i\xi} \\
i\tau e^{-i\xi} & \rho e^{-i\eta}
\end{array}
\right), \ \tau\in[-1,1], \ \rho=\sqrt{1-\tau^2}, 
\end{equation}
where $\psi$, $\eta$, and $\xi$ are real numbers. The matrix is dependent on four parameters
but, as we shall see, three of them can be removed by redefining the scattering channels. Care is needed, though,
since redefining the channels one should preserve the time-reversal symmetry of the channels. Let us apply a diagonal transformation matrix
\begin{equation}\label{Udiag}
\widehat{U}={\rm diag}(e^{i\psi_1},e^{i\psi_2},e^{i\psi_3},e^{i\psi_4})
\end{equation}
to the incident channels so that
\begin{equation}
\tilde{\bf s}_{\smallp}=\widehat{U}{\bf s}_{\smallp},
\end{equation}
then the outgoing channels have to be transformed as 
\begin{equation}
\tilde{\bf s}_{\smallm}=\widehat{U}^*{\bf s}_{\smallm}.
\end{equation}
By substituting the unitary transformation into Eq.~\eqref{0A3} one finds that the choice
\begin{align}
  \psi_1=\frac{1}{2}(\psi+\eta+\xi),      \nonumber \\
   \psi_2=\frac{1}{2}(\psi-\eta-\xi),      \nonumber \\
    \psi_3=\frac{1}{2}(\psi+\eta-\xi),      \nonumber \\
     \psi_4=\frac{1}{2}(\psi-\eta+\xi)
\end{align}
leads to
\begin{equation}
\widehat{S}=
\left(
\begin{array}{cc}
\rho  & i\tau \\
i\tau  & \rho
\end{array}\label{canonical}
\right), \ \tau\in[-1,1], \ \rho=\sqrt{1-\tau^2}.
\end{equation}
Thus, $\widehat{S}$ can be written in the canonical form of a symmetric unitary matrix. It
is important to note that $\widehat{S}_{\scs{4\!\times\!4}}$ is frequency dependent. Therefore,
albeit the structure of the matrix given by Eq.~\eqref{0A3} along with the unitarity of $\widehat{S}$ is preserved in a frequency sweep, the transformation $\widehat{U}$ is also frequency dependent. More precisely, one can guaranty Eq.~\eqref{canonical} only in a single point in a frequency domain with a frequency independent $\widehat{U}$. Note that Eq.~\eqref{canonical} is identical to that proposed in \cite{Zhou2016} up to an arbitrary phase factor.

%%%%%%%%%%%%%%%%%%%%%%%%%%%%%%%%%%%%%%%%%%%%%%%%%%%%%%%%%%%%%%%%%%%%%%%
%%%%%%%%%%%%%%%%%%%%%%%%%%%%%%%%%%%%%%%%%%%%%%%%%%%%%%%%%%%%%%%%%%%%%%%%%%%%

\section{TCMT revisited} \label{Sec3}
The key idea behind the TCMT is to cast the scattering problem in the form of a system of coupled driven oscillators. Each oscillator in the system represents an optical resonant mode which can be excited by incident light.
Mathematically, in the TCMT framework the dynamics is controlled by a system of linear ordinary differential equations.
The credit of the modern TCMT formulation is to Fan, Suh, and Joannopoulos who were the first to have put forward a number of important relationships for the system's coefficients rendering the TCMT tractable for describing Fano resonances in a single mode approach
\cite{Fan03}. Later, in~\cite{Suh04} the TCMT was generalized to the multi-modal case. Generalizations for arbitrary-shaped objects with an infinite number of scattering channels are also possible \cite{Ruan12, Bulgakov2023}.
More recently, the effects of the time-reversal symmetry, energy conservation, and reciprocity on the TCMT constraints were thoroughly investigated in~\cite{zhao2019connection}. 

The TCMT is akin to the plethora of resonant state expansion methods which have recently gained popularity in the photonics community \cite{muljarov2018resonant, lalanne2018light, sauvan2022normalization}. These methods also rely on series expansions into eigenmodes of Maxwell's equation with radiation boundary conditions. The difference is, though, that the TCMT, being a phenomenological approach, does not utilize the explicit form of the eigenmodes, but rather establishes the generic form of the optical response relying on physical constraints such as symmetries and conservation laws. Practically, the TCMT solutions always contain unknown coefficients which are to be fitted to comply with exact full-wave solutions obtained by numerically exact methods. In this work we use the FDFD method~\cite{rumpf2022electromagnetic} (see Sec.~\ref{Sec5}).  Yet, the choice of simulation method is, of course, always at the discretion of the researcher, so other methods such as the increasingly popular finite-difference time-domain and finite-element methods could have been equally applied. 
Finally, it is worth mentioning that recently some attempts have been made to put the TCMT on more rigorous grounds by deriving the TCMT equations from resonant mode expansions
\cite{Alpeggiani17, zhang2020quasinormal, Wu2024}.

However, it is worth mentioning that recently some attempts have been made to put the TCMT on more rigorous grounds by deriving the TCMT equations from resonant mode expansions
\cite{Alpeggiani17, zhang2020quasinormal, Wu2024}.

According to~\cite{Suh04} the TCMT the relationship between ${\bf s}_{\smallp}$ and ${\bf s}_{\smallm}$ can be written in the following form
\begin{align}\label{TCMT}
    & \frac{d{\bf a}(t)}{dt}=[-i\widehat{\Omega}-\widehat{\Gamma}]{\bf a}(t)+\widehat{K}^{\intercal}
    {\bf s}_{\smallp}(t)\nonumber, \\
    & {\bf s}_{\smallm}(t) = 
    \widehat{C}{\bf s}_{\smallp}(t)+\widehat{D}{\bf a}(t),
 \end{align}
where ${\bf a}(t)$ is a column vector composed of the amplitudes of the resonant eigenmodes supported by the photonic structure, 
%${\bf a}(t)=[a_1(t),  a_2(t), \ldots,a_M(t)]^{\intercal}$, 
$\widehat{C}$ is the matrix of direct (non-resonant) process, $\widehat{\Omega}$ is the Hermitian matrix containing the natural frequencies of the resonant modes,  $\widehat{\Gamma}$ is the Hermitian matrix of decay rates, $\widehat{K}$ is the coupling matrix, and $\widehat{D}$ is the decoupling matrix. In the frequency domain, i.e.~for systems illuminated by monochromatic light, we apply a time-harmonic substitution for all
time-dependent quantities, e.g.
\begin{equation}
{\bf a}(t)={\bf a}e^{-i\omega t},
\end{equation}
which leads to 
\begin{align}\label{TCMT_freq}
    & [i(\widehat{\Omega}-\omega\widehat{\mathbb I})+\widehat{\Gamma}]{\bf a}=\widehat{K}^{\intercal}
    {\bf s}_{\smallp}\nonumber, \\
    & {\bf s}_{\smallm} = 
    \widehat{C}{\bf s}_{\smallp}+\widehat{D}{\bf a}.
\end{align}
Equations~\eqref{TCMT_freq} are the basic equations
to be used for finding the $S$-matrix.

Now, for the sake of generality, assume that there are $M$ resonant modes coupled with $N$ scattering channels. 
%The time-harmonic TCMT equations can be written as
%\begin{align}\label{0Amany}
 %   & [i(\widehat{\Omega}-\omega\widehat{\mathbb %I})+\widehat{\Gamma}]{\bf a}=\widehat{D}^{\intercal}
%    {\bf s}^{\smallp}\nonumber, \\
 %   & {\bf s}^{\smallm} = 
%    \widehat{C}{\bf s}^{\smallp}+\widehat{D}{\bf a},
%\end{align}
Then the quantities in Eq.~\eqref{TCMT} are as follows. The vector ${\bf a}$ is a $M\!\times\!1$ column vector, $\widehat{D}$ and $\widehat{K}$ are $N\!\times\!M$ matrices, and $\widehat{C}$ is a $N\!\times\!N$ matrix. In the same manner as in \cite{Suh04}, by taking into account the photon number conservation in radiation loss from an unpumped system  one can show that Eq.~\eqref{TCMT} leads to
\begin{equation}\label{0Agamma}
2\widehat{\Gamma}=\widehat{D}^{\dagger}\widehat{D}.
\end{equation}
At the same time the general solution for the $S$-matrix can be obtained from Eq.~\eqref{TCMT_freq} as follows
\begin{equation}\label{0ASmany}
\widehat{S}(\omega)=\widehat{C}+\widehat{D}\frac{1}{i(\widehat{\Omega}-\omega\widehat{\mathbb I})+\widehat{\Gamma}}\widehat{K}^{\intercal}.
\end{equation}
The unitarity of the $S$-matrix given by Eq.~\eqref{0ASmany} leads to a rather awkward equation
\begin{align}\label{0A13}
& \widehat{K}^*\frac{1}{-i(\widehat{\Omega}-\omega\widehat{\mathbb I})+\widehat{\Gamma}}\widehat{D}^{\dagger}\widehat{C}+
\widehat{C}^*\widehat{D}\frac{1}{i(\widehat{\Omega}-\omega\widehat{\mathbb I})+\widehat{\Gamma}}\widehat{K}^{\intercal} \nonumber \\
&
+2\widehat{K}^*\frac{1}{-i(\widehat{\Omega}-\omega\widehat{\mathbb I}) +\widehat{\Gamma}}\widehat{\Gamma}\frac{1}{i(\widehat{\Omega}-\omega\widehat{\mathbb I})+\widehat{\Gamma}}\widehat{K}^{\intercal}=0.
\end{align}
To make use out of Eq.~\eqref{0A13} we consider the eigenvalue problem
\begin{equation}\label{0A14}
(i\widehat{\Omega}+\widehat{\Gamma}){\bf a}_m=i\omega_m{\bf a}_m.
\end{equation}
In what follows we focus on the generic case when there exist $M$ linearly independent, though not necessarily orthogonal, eigenvectors. In the other words, it is assumed that there are no exceptional points \cite{Oezdemir19} in the spectrum. The vectors ${\bf a}_m$ will be termed the right eigenvectors of $i\widehat{\Omega}+\widehat{\Gamma}$. The right eigenvectors can be complemented with the left eigenvectors ${\bf b}_m$ which solve the eigenvalue problem
\begin{equation}\label{0A15}
{\bf b}_m^{\intercal}(i\widehat{\Omega}+\widehat{\Gamma})=i\omega_m{\bf b}_m^{\intercal}.
\end{equation}
It is possible to show that the right and left eigenvectors can be normalized such that the following orthonormalization condition holds true
\begin{equation}\label{0A16}
{\bf b}_{m}^{\intercal}{\bf a}_{m'}=\delta_{m,m'}.
\end{equation}
Note that ${\bf a}_m$ are not orthogonal in terms of the skew-symmetric inner product ${\bf a}_{m}^{\dagger}{\bf a}_{m'}\neq\delta_{m,m'}$. Thus, one would rather
say that ${\bf a}_m$ and ${\bf b}_m$ form a bi-orthogonal basis \cite{Rotter2009}. Since the vector space dimensionality is finite one can show
that the biorthoganal vectors resolve the identity operator as follows
\begin{equation}\label{0A17}
\widehat{\mathbb I}=\sum_{m=1}^M{\bf a}_{m}{\bf b}_{m}^{\intercal}.
\end{equation}
The matrix in the TCMT equations Eq.~\eqref{TCMT_freq} can be written in terms of the left and right eigenvectors
\begin{equation}\label{0A18}
i(\widehat{\Omega}-\omega\widehat{\mathbb I})+\widehat{\Gamma}=
i\sum_{m=1}^M(\omega_m-\omega){\bf a}_{m}{\bf b}_{m}^{\intercal}.    
\end{equation}
The following useful relationships can be derived from Eq.~\eqref{0A17} and Eq.~\eqref{0A18}
\begin{equation}\label{0A19}
\frac{1}{i(\widehat{\Omega}-\omega\widehat{\mathbb I})+\widehat{\Gamma}}=-i
\sum_{m=1}^M\frac{1}{\omega_m-\omega}{\bf a}_{m}{\bf b}_{m}^{\intercal},
\end{equation}
and  
\begin{equation}\label{0A24}
2\widehat{\Gamma}=i\sum_{m=1}^M(\omega_m-\omega){\bf a}_{m}{\bf b}_{m}^{\intercal}
-i\sum_{m=1}^M(\omega_m^*-\omega){\bf b}^*_{m}{\bf a}_{m}^{\dagger}.
\end{equation}
%For the future purpose we also introduce auxiliary functions
%\begin{align}\label{0A20}
%& \Pi_{\scs{M,0}}(\omega)=i^{\scs{M}}\prod_{m=1}^M(\omega_m-\omega),
%\end{align}
%and
%\begin{align}\label{0A21}
%& \Pi_{\scs{M,m}}(\omega)=i^{\scs{M-1}}\prod_{m'\neq m}^M(\omega_{m'}-\omega). 
%\end{align}

Now we are well equipped for analyzing Eq.~\eqref{0A13}. By plugging Eq.~\eqref{0A19} into Eq.~\eqref{0A13} one finds 
\begin{align}\label{0A22}
& i\sum_{m=1}^{M}\widehat{C}^*\widehat{D}\frac{{\bf a}_m{\bf b}_m^{\intercal}}{\omega_m-\omega}\widehat{K}^{\intercal}-i
\sum_{m=1}^{M}\widehat{K}^*\frac{{\bf b}_m^*{\bf a}_m^{\dagger}}{\omega_m^*-\omega}\widehat{D}^{\dagger}\widehat{C}
= \nonumber \\
& 2\sum_{m,m'=1}^{M}
\widehat{K}^{*}\frac{{\bf b}_m^*{\bf a}_m^{\dagger}}{\omega_m^{*}-\omega} \widehat{\Gamma}
\frac{{\bf a}_{m'}{\bf b}_{m'}^{\intercal}}{\omega_{m'}-\omega}\widehat{K}^{\intercal},
\end{align}
%The leading term in the above polynomial expression in $\omega$ is of order $M(M+1)$. By %applying the resolution of identity, Eq.~\eqref{0A19} in the leading term one obtains
%\begin{equation}\label{0A23}
%\widehat{K}^*\widehat{D}^{\dagger}\widehat{C}=\widehat{C}^*\widehat{D}\widehat{K}^{\intercal%}.
%\end{equation}
Substituting Eq.~\eqref{0A24} into Eq.~\eqref{0A22} and collecting the terms with the same poles in the $\omega$-plane one obtains
\begin{equation}\label{0A25}
\widehat{K}^*{\bf b}_m^*{\bf a}_m^{\dagger}(\widehat{D}^{\dagger}\widehat{C}+\widehat{K}^{\intercal})=0.
\end{equation}
After summation over $m$ and application of Eq.~\eqref{0A17} we have
\begin{equation}\label{0A26}
\widehat{K}^*(\widehat{D}^{\dagger}\widehat{C}+\widehat{K}^{\intercal})=0.
\end{equation}
In the same manner, it can be shown that the symmetry of the $S$-matrix leads to
\begin{equation}\label{0A27}
\widehat{D}\widehat{K}^{\intercal}=\widehat{K}\widehat{D}^{\intercal}.
\end{equation}
It is worth mentioning that the above equation may also hold for lossy systems, see the supplemental material to \cite{Zhou2016}, since it can be derived solely relying on the Lorentz reciprocity.

Let us now analyze our findings. First of all we recall that according to \cite{Suh04} the TMCT matching condition in the presence of energy conservation and time-reversal symmetry have the following form
\begin{align}\label{0A28}
& \widehat{C}\widehat{D}^*+\widehat{D}=0, \nonumber \\
&\widehat{D}=\widehat{K}.
\end{align}
One can easily see that if the matrices $\widehat{D}$ and $\widehat{K}$ satisfy Eq.~\eqref{0A28} then they also solve the new found Eq.~\eqref{0A26} and Eq.~\eqref{0A27}.
One can argue, however, that there are solutions of Eq.~\eqref{0A26} and Eq.~\eqref{0A27} not accommodated in Eq.~\eqref{0A28}. 

All scattering observables cannot depend on how we label the resonant states and the theory must remain unchanged under an arbitrary unitary transformation of the basis. Therefore, all relations between $\widehat{C}$, $\widehat{D}$, and $\widehat{K}$ must be {\it basis-invariant}. Let us check this and consider a unitary transformation in space of the resonator eigenmodes ${\bf a}\rightarrow{\bf a}'$, bearing in mind that unitarity is necessary for maintaining Eq.~\eqref{0Agamma}. This unitary transformation can be written as
\begin{equation}\label{0A29}
{\bf a}=\widehat{U}{\bf a}',
\end{equation}
where $\widehat{U}$ is a unitary matrix.
By plugging Eq.~\eqref{0A29} into the TCMT equations Eq.~\eqref{TCMT} we find that the matrix quantities are transformed as follows
\begin{align}\label{0A30}
& i\widehat{\Omega}'+\widehat{\Gamma}'=\widehat{U}^{\dagger}(i\widehat{\Omega}+\widehat{\Gamma})\widehat{U}, \nonumber \\
& \widehat{D}'=\widehat{D}\widehat{U}, \nonumber \\
& \widehat{K}'=\widehat{K}\widehat{U}^{*}.
\end{align}
One can easily check that the TCMT matching conditions given by Eq.~\eqref{0Agamma},  Eq.~\eqref{0A26}, and Eq.~\eqref{0A27} are invariant to any unitary transformation whereas Eq.~\eqref{0A28} is not. This is a consequence of that the matrix $\widehat{K}$ is covariant with ${\bf a}$ while $\widehat{D}$ is contra-variant. The $\mathrm{U}(M)$ invariant matching conditions that we have derived in this section obviously lead to new solutions not accommodated in Eq.~\eqref{0A28} after some unitary transformations.

%%%%%%%%%%%%%%%%%%%%%%%%%%%%%%%%%%%%%%%%%%%%%%
%%%%%%%%%%%%%%%%%%%%%%%%%%%%%%%%%%%%%%%%%%%%%%%%%
\section{TCMT for grating} \label{Sec4}
\subsection{Two contra-propagating resonance modes}

Let us now apply the results of the previous section to the system introduced in Sec.~\ref{Sec2}.
The $4\!\times\!4$ $S$-matrix, which was defined in Eq.~\eqref{Smatrix} can be rewritten in a more convenient form
\begin{equation}
\left(\begin{array}{c}
{\bf s}_{\sml}^{\smallm} \\
{\bf s}_{\smr}^{\smallm}
\end{array}\right)=
\left(
\begin{array}{cc}
0 & \widehat{S} \\
\widehat{S}^{\intercal} & 0
\end{array}
\right)
\left(
\begin{array}{c}
{\bf s}_{\smr}^{\smallp} \\
{\bf s}_{\sml}^{\smallp}
\end{array}
\right),
\end{equation}
where
% \begin{equation}
% {\bf s}_{\smr,(\sml)}^{\smallpm}=
% \left(\begin{array}{c}
% { s}_{1}^{\smallpm} \\
% { s}_{2}^{\smallpm}
% \end{array}
% \right), \
% {\bf s}_{\sml,(\smr)}^{\smallpm}=
% \left(\begin{array}{c}
% { s}_{3}^{\smallpm} \\
% { s}_{4}^{\smallpm}
% \end{array}
% \right).
% \end{equation}
\begin{equation}
{\bf s}_{\smr}^{\smallp}, {\bf s}_{\sml}^{\smallm}=
\left(\begin{array}{c}
{ s}_{1}^{\smallpm} \\
{ s}_{2}^{\smallpm}
\end{array}
\right), \
{\bf s}_{\sml}^{\smallp}, {\bf s}_{\smr}^{\smallm}=
\left(\begin{array}{c}
{ s}_{3}^{\smallpm} \\
{ s}_{4}^{\smallpm}
\end{array}
\right).
\end{equation}
Note that the orientation of the subscript arrows in the above equations corresponds to the wave propagation direction with respect to the $x$-axis according to Fig.~\ref{Fig2}. Thus, taking into account the momentum conservation \cite{Zhou2016} the coupling and decoupling matrices can also be written in a simplified form as follows
\begin{equation}
\widehat{K}=
\left(
\begin{array}{cc}
0 & {\bf \kappa}_{\smr} \\
 {\bf \kappa}_{\sml} & 0
\end{array}
\right), \ {\bf \kappa}_{\smr}=
\left(
\begin{array}{c}
\varkappa_1 \\
\varkappa_2
\end{array}
\right), \ {\bf \kappa}_{\sml}=
\left(
\begin{array}{c}
\varkappa_3 \\
\varkappa_4
\end{array}
\right),
\end{equation}
and
\begin{equation}\label{Dmat}
\widehat{D}=
\left(
\begin{array}{cc}
{\bf d}_{\sml} & 0 \\
 0& {\bf d}_{\smr}
\end{array}
\right), \ {\bf d}_{\smr}=
\left(
\begin{array}{c}
d_3 \\
d_4
\end{array}
\right), \ {\bf d}_{\sml}=
\left(
\begin{array}{c}
d_1 \\
d_2
\end{array}
\right),
\end{equation}
where the subscript enumerates the scattering channels according to Fig.~\ref{Fig2}. 
The coupling and decoupling matrices are of a block form in compliance with the momentum conservation. In accordance with our definition of $\widehat{D}$ and $\widehat{K}$, the time-domain TCMT equations~\eqref{TCMT} take the following form  
%\begin{equation}
%\widehat{K}=
%\left(
%\begin{array}{cc}
%0 & \kappa_1 \\
%0 & \kappa_2 \\
%\kappa_3 & 0 \\
%\kappa_4 & 0
%\end{array}
%\right)
%\end{equation}
%\begin{equation}
%\widehat{D}=
%\left(
%\begin{array}{cc}
%d_1 & 0 \\
%d_2 & 0 \\
%0 & d_3 \\
%0 & d_4
%\end{array}
%\right)
%\end{equation}
\begin{align}
& \frac{d}{dt}\left(
\begin{array}{c}
a_{\sml} \\
a_{\smr}
\end{array}
\right)=
\left(
\begin{array}{cc}
-i\omega_0-\gamma & 0 \\
0 & -i\omega_0-\gamma
\end{array}
\right)
\left(
\begin{array}{c}
a_{\sml} \\
a_{\smr}
\end{array}
\right)+\widehat{K}^{\intercal}
\left(
\begin{array}{c}
{\bf s}_{\smr}^{\smallp} \\
{\bf s}_{\sml}^{\smallp}
\end{array}
\right), \nonumber \\
&
\left(\begin{array}{c}
{\bf s}_{\sml}^{\smallm} \\
{\bf s}_{\smr}^{\smallm}
\end{array}
\right)=\widehat{C}_{\scs{4\!\times\!4}}
\left(
\begin{array}{c}
{\bf s}_{\smr}^{\smallp} \\
{\bf s}_{\sml}^{\smallp}
\end{array}
\right)+\widehat{D}\left(
\begin{array}{c}
a_{\sml} \\
a_{\smr}
\end{array}
\right).
\end{align}

In the next step we apply the TCMT matching equations Eq.~\eqref{0Agamma}, ~Eq.~\eqref{0A26}, and ~Eq.~\eqref{0A27}. Equation~\eqref{0A27} yields
\begin{equation}
\widehat{D}=
A\left(
\begin{array}{cc}
{\bf \kappa}_{\smr} & 0 \\
0 & {\bf \kappa}_{\sml}
\end{array}
\right),
\end{equation}
where $A$ is an arbitrary complex number. Substituting the above into
$\widehat{D}^{\dagger}\widehat{D}=2\widehat{\Gamma}$ gives
\begin{equation}
A=e^{i\zeta}\sqrt{\frac{2\gamma}{\kappa^2}},
\end{equation}
where $\zeta\in[0,~2\pi)$ is a real number and
\begin{equation}
\kappa^2={\bf \kappa}_{\sml}^{\dagger}{\bf \kappa}_{\sml}={\bf \kappa}_{\smr}^{\dagger}{\bf \kappa}_{\smr}
\end{equation}
so the vectors ${\bf \kappa}_{\sml}$ and ${\bf \kappa}_{\smr}$ are equal in the norm. Before applying Eq.~\eqref{0A26} we recall that according to Eq.~\eqref{0A3} and Eq.~\eqref{canonical} the $4\!\times\!4$ $S$-matrix can be written in a block form with a symmetric unitary matrix off the main diagonal at any given frequency. Here we apply this result to the matrix of the direct process which, being independent of frequency, can be written as follows
\begin{equation}\label{C44}
\widehat{C}_{\scs{4\!\times\!4}}=
\left(
\begin{array}{cc}
0 & \widehat{C} \\
 \widehat{C}& 0
\end{array}
\right),
\end{equation}
where
\begin{equation}
\widehat{C}=
\left(
\begin{array}{cc}
\rho & i\tau \\
i\tau& \rho
\end{array}
\right).
\end{equation}

Note, however, that once the unitary transformation in Eq.~\eqref{Udiag} is applied, thereby bringing $\widehat{C}_{4\times 4}$ to the diagonal form of Eq.~\eqref{C44}, the channel functions are no longer mapped directly onto one another. As a result, every non-zero entry in the matrix representation of a symmetry operation acquires a phase factor determined by $U_{\text{diag}}$. Consequently, symmetry constraints can be enforced only by matching the \emph{magnitudes} of the coupling-vector elements, not their complex phases.

After using Eq.~\eqref{0A26} and assuming in accordance with $\gamma>0$ that both vectors ${\bf \kappa}_{\sml}$ and  ${\bf \kappa}_{\smr}$ have at least one non-zero element, we obtain
\begin{align}\label{kappa_eqn}
e^{-i\zeta}\sqrt{\frac{2\gamma}{\kappa^2}}\widehat{C}{\bf \kappa}_{\smr}^{*}+{\bf \kappa}_{\sml}=0, \nonumber \\
e^{-i\zeta}\sqrt{\frac{2\gamma}{\kappa^2}}\widehat{C}{\bf \kappa}_{\sml}^{*}+{\bf \kappa}_{\smr}=0.
\end{align}
The solution of the above equation is any vector $\kappa_{\smr} (\kappa_{\sml})$ that satisfies the normalization condition
\begin{equation}
\kappa^2=2\gamma.
\end{equation}
Now, our goal is to write such a solution in a form symmetric with respect to both $\kappa_{\smr}$ and $\kappa_{\sml}$. To do this
we  introduce matrix $\widehat{B}$ that is a symmetric unitary square root of $-\widehat{C}$ so that
\begin{align}\label{B}
& \widehat{B}\widehat{B}=-\widehat{C}, \notag \\
& \widehat{B}^{\intercal}=\widehat{B}, \notag \\
& \widehat{B}\widehat{B}^{\dagger}=\widehat{\mathbb I}.
\end{align}
Matrix $\widehat{B}$ can be explicitly written as
\begin{equation}
\widehat{B}=
{\frac{1}{\sqrt{2(1+\rho)}}}
\left(
\begin{array}{cc}
\tau & -i(1+\rho) \\
-i(1+\rho) & \tau
\end{array}
\right).
\end{equation}
Having in mind Eq.~\eqref{B}, the solution of Eq.~\eqref{kappa_eqn} can be written as
\begin{align}\label{kappa_main}
& {\bf \kappa}_{\smr}=\sqrt{2\gamma}e^{-i\zeta/2}\widehat{B}{\bf n}^{*}, \nonumber \\ 
& {\bf \kappa}_{\sml}=\sqrt{2\gamma}e^{-i\zeta/2}\widehat{B}{\bf n},
\end{align}
where $\bf n$ is any $2\times 1$ vector such as ${\bf n}^{\dagger}{\bf n}=1$.
The $S$-matrix for the right going waves is then written as
\begin{equation}\label{right-going}
\widehat{S}=\widehat{C}+\frac{2\gamma}{i(\omega_0-\omega)+\gamma}\widehat{B}{\bf n}{\bf n}^{\dagger}\widehat{B}.
\end{equation}
%while for the left-going waves we have 
%\begin{equation}
%\widehat{S}=\widehat{C}+\frac{2\gamma}{i(\omega_0-%\omega)+\gamma}\widehat{B}{\bf n}^*{\bf %n}^{\intercal}\widehat{B},
%\end{equation}
For the left-going waves the $S$-matrix is the transpose of Eq.~\eqref{right-going} according to Eq.~\eqref{0A3}. 
Using Eq.~\eqref{B} it is possible to demonstrate that $\widehat{S}$ is a unitary matrix. 
%The solution for the $S$-matrix is independent of $\eta$ which can be dropped in further calculations.

The unit vector $\bf n$ can be parametrized by three variables
\begin{align}\label{n_main}
& {\bf n}=
e^{i\chi}\left(
\begin{array}{c}
\cos\alpha~e^{+i\delta/2} \\
\sin\alpha~e^{-i\delta/2}
\end{array}
\right), \nonumber \\
& \chi\in[0, 2\pi), \ \alpha\in[-\pi/2, \pi/2], \ \delta\in[0, \pi),
\end{align}
which leads to the following solution 
%%%%%%%%%%%%%%%%%%%%%%%%%%%%%%%%%%%%%%%%%%%%%%%%%%
% for the coupling 
% \begin{align}\label{kleft}
% {\bf \kappa}_{\smr}=
% e^{-i\psi-i\eta/2}\sqrt{\frac{\gamma}{1+\rho}}
% \left(
% \begin{array}{c}
% \tau\cos\alpha~e^{-i\delta/2} -i(1+\rho)\sin\alpha~e^{+i\delta/2} \\
% -i(1+\rho)\cos\alpha~e^{-i\delta/2}+\tau\sin\alpha~e^{+i\delta/2}
% \end{array}
% \right)
% \end{align}
% and the decoupling
% \begin{align}\label{dleft}
% {\bf d}_{\smr}=
% e^{i\psi+i\eta/2}\sqrt{\frac{\gamma}{1+\rho}}
% \left(
% \begin{array}{c}
% \tau\cos\alpha~e^{+i\delta/2} -i(1+\rho)\sin\alpha~e^{-i\delta/2} \\
% -i(1+\rho)\cos\alpha~e^{+i\delta/2}+\tau\sin\alpha~e^{-i\delta/2}
% \end{array}
% \right)
% \end{align}
% vectors. In the same manner for the left going waves we can write
% \begin{equation}\label{kright}
% {\bf \kappa}_{\sml}=
% e^{+i\psi-i\eta/2}\sqrt{\frac{\gamma}{1+\rho}}
% \left(
% \begin{array}{c}
% \tau\cos\alpha~e^{+i\delta/2} -i(1+\rho)\sin\alpha~e^{-i\delta/2} \\
% -i(1+\rho)\cos\alpha~e^{+i\delta/2}+\tau\sin\alpha~e^{-i\delta/2}
% \end{array}
% \right),
% \end{equation}
% and
% \begin{equation}\label{dright}
% {\bf d}_{\sml}=
% e^{-i\psi+i\eta/2}\sqrt{\frac{\gamma}{1+\rho}}
% \left(
% \begin{array}{c}
% \tau\cos\alpha~e^{-i\delta/2} -i(1+\rho)\sin\alpha~e^{+i\delta/2} \\
% -i(1+\rho)\cos\alpha~e^{-i\delta/2}+\tau\sin\alpha~e^{+i\delta/2}
% \end{array}
% \right)
% \end{equation}
%%%%%%%%%%%%%%%%%%%%%%%%%%%%%%%%%%%%%%%%%%%
for the transmittance
\begin{align}\label{TT}
T=\frac{[\tau(\omega_0-\omega)+ \rho\gamma\sin(2\alpha)\cos\delta]^2+\gamma^2\sin^2 (2\alpha)\sin^2\delta}
{(\omega_0-\omega)^2+\gamma^2}.
\end{align}
Note that the above result is independent of the choice of the incidence channel. In contrast, the analytic expressions for the squared resonant mode amplitude are different depending on the direction of incidence. When the metasurface is illuminated through channel $s_1^{\smallp}$ the solution reads
\begin{equation}\label{a14}
|a_{\smr}|^2=\frac{\gamma[1-\rho\cos(2\alpha)+\tau\sin(2\alpha)\sin\delta]}
{(\omega_0-\omega)^2+\gamma^2},
\end{equation}
whereas with channel $s_3^{\smallp}$ the solution is 
\begin{equation}\label{a23}
|a_{\sml}|^2=\frac{\gamma[1-\rho\cos(2\alpha)-\tau\sin(2\alpha)\sin\delta]}
{(\omega_0-\omega)^2+\gamma^2}.
\end{equation}
The solution for the $S$-matrix Eq.~\eqref{right-going} does not depend on the phases $\chi$ and $\zeta$. Further on we set
%Although the coupling and decoupling vectors can be %assigned arbitrary $\psi$ care is needed when %considering normal incidence. At normal incidence %the left- and the right-going (de)coupling vectors %must be identical. This results in the only %possible choice
$ \chi=0, \ \zeta=0,$
which gives 
\begin{equation}\label{coup_decoup}
{\bf d}_{\sml}={\bf \kappa}_{\smr}, \ {\bf d}_{\smr}={\bf \kappa}_{\sml}.
\end{equation}
The expression for the coupling vectors can be written as
\begin{align}\label{kappa_final}
{\bf \kappa}_{\smr,\sml}=\!
\!\sqrt{\frac{\gamma}{1+\rho}}
\!\left(\!
\begin{array}{c}
\tau\cos\alpha~e^{\mp i\delta/2} -i(1+\rho)\sin\alpha~e^{\pm i\delta/2} \\
-i(1+\rho)\cos\alpha~e^{\mp i\delta/2}+\tau\sin\alpha~e^{\pm i\delta/2}
\end{array}
\!\right)\! .
\end{align}
%%%%%%%%%%%%%%%%%%%%%%%%%%%%%%%%%%%%%%%%%
%%%%%%%%%%%%%%%%%%%%%%%%%%%%%%%%%%%%%%%%%%%%%%%%%%%%%%%%%%%%%%%%%%%%%%%%%%%%%%%%%%%%

\subsection{Fano resonance}

Since the Fano resonance is a product of interference between a resonant and a non-resonant pathways~\cite{Limonov17}, the 
following generic expression holds for the transmittance coefficients
\begin{equation}
\label{T_Fano}
T =  \left|A_1 + \frac{A_2 e^{i\Delta}}{\Omega + i}\right|^2,
\end{equation}
where $A_1, A_2, \Delta_2$ are real numbers, and $\Omega = (\omega - \omega_0)/\gamma$. 
Equation~\eqref{T_Fano}, besides $\Omega$, depends on three real-valued parameters, same as Eq.~\eqref{TT}. It is possible to
show after some algebra that the parameters in Eq.~\eqref{TT} and Eq.~\eqref{T_Fano} are linked by the following formulas
\begin{align}
\label{TCMT_vs_Fano}
A_1 & = |\tau|, \\
A_2 & = \sqrt{(\tau + \sin (2\alpha)\sin\delta)^2 + \rho^2\sin^2(2\alpha)\cos^2\delta}, \\
\Delta & =  - \arctan{\left(\frac{\rho\sin (2\alpha)\cos\delta}{\tau + \sin (2\alpha)\sin\delta}\right)} -\operatorname{sign}\left(\tau\right)\frac{\pi}{2}.
\end{align}
Alternatively, according to \cite{Limonov17}, the Fano formula~\eqref{T_Fano} can be also expressed as  
\begin{figure}[t]
    \centering    %\includegraphics[width=1\linewidth]%{Fig_2.png}
    \includegraphics[width=1\linewidth]{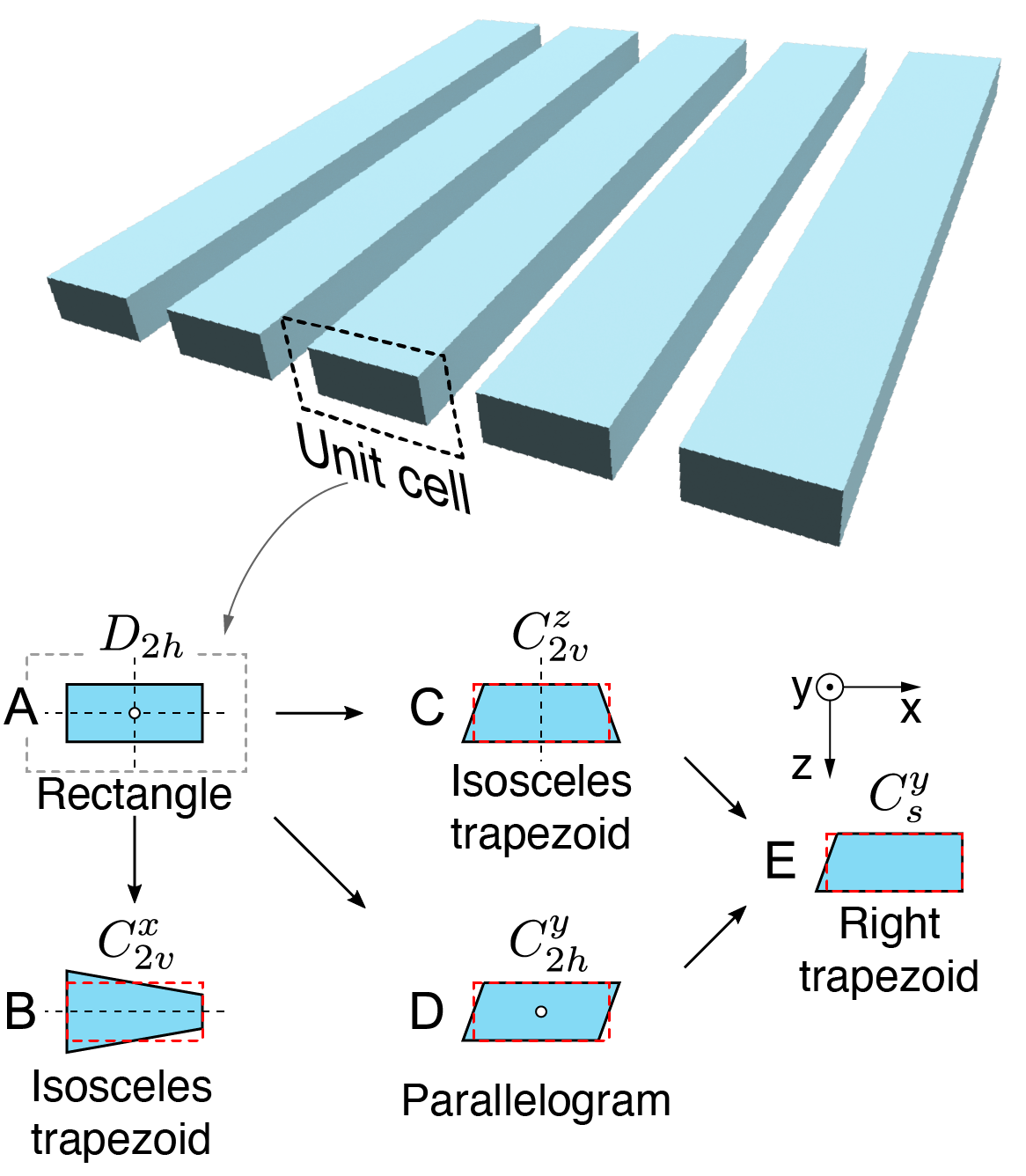} \caption{Grating with a unit cell of various symmetries. A -- rectangular unit cell (D$_{2h}$), B -- isosceles trapezoid unit cell (C$_{2v}^x$), C -- isosceles trapezoid unit cell (C$_{2v}^z$), D -- parallelogram unit cell (C$_{2h}^y$), E -- right trapezoid unit cell (C$_{s}^y$).}    \label{Fig3}
\end{figure} 
\begin{equation}
\label{T_Rybin}
T = \left[\frac{(q+\Omega)^2}{1+\Omega^2} p + (1-p)\right] A_1^2,
\end{equation}
where $p \in[0, 1]$ is the interaction coefficient~\cite{Rybin2013} 
\begin{equation}
p=\frac{2 F \cos ^2 \Delta}{F+2 \sin \Delta+\sqrt{F^2+4 F \sin \Delta+4}},
\end{equation}
$F = A_2/A_1$ is the relative intensity, and $q$ is the Fano asymmetry parameter,
\begin{equation}\label{asymmetry}
q = \frac{F \cos\Delta}{p}.
\end{equation}
Thus, Eq.~\eqref{TT} complies with the generic theories of the Fano.
%%%%%%%%%%%%%%%%%%%%%%%%%%%%%%%%%%%%%%%%%%%%%%%
%%%%%%%%%%%%%%%%%%%%%%%%%%%%%%%%%%%%%%%%%%%%%%%
%%%%%%%%%%%%%%%%%%%%%%%%%%%%%%%%%%%%%%%%%%%%%%%

\begin{figure*}[th]
    \centering
    \includegraphics[width=\linewidth]{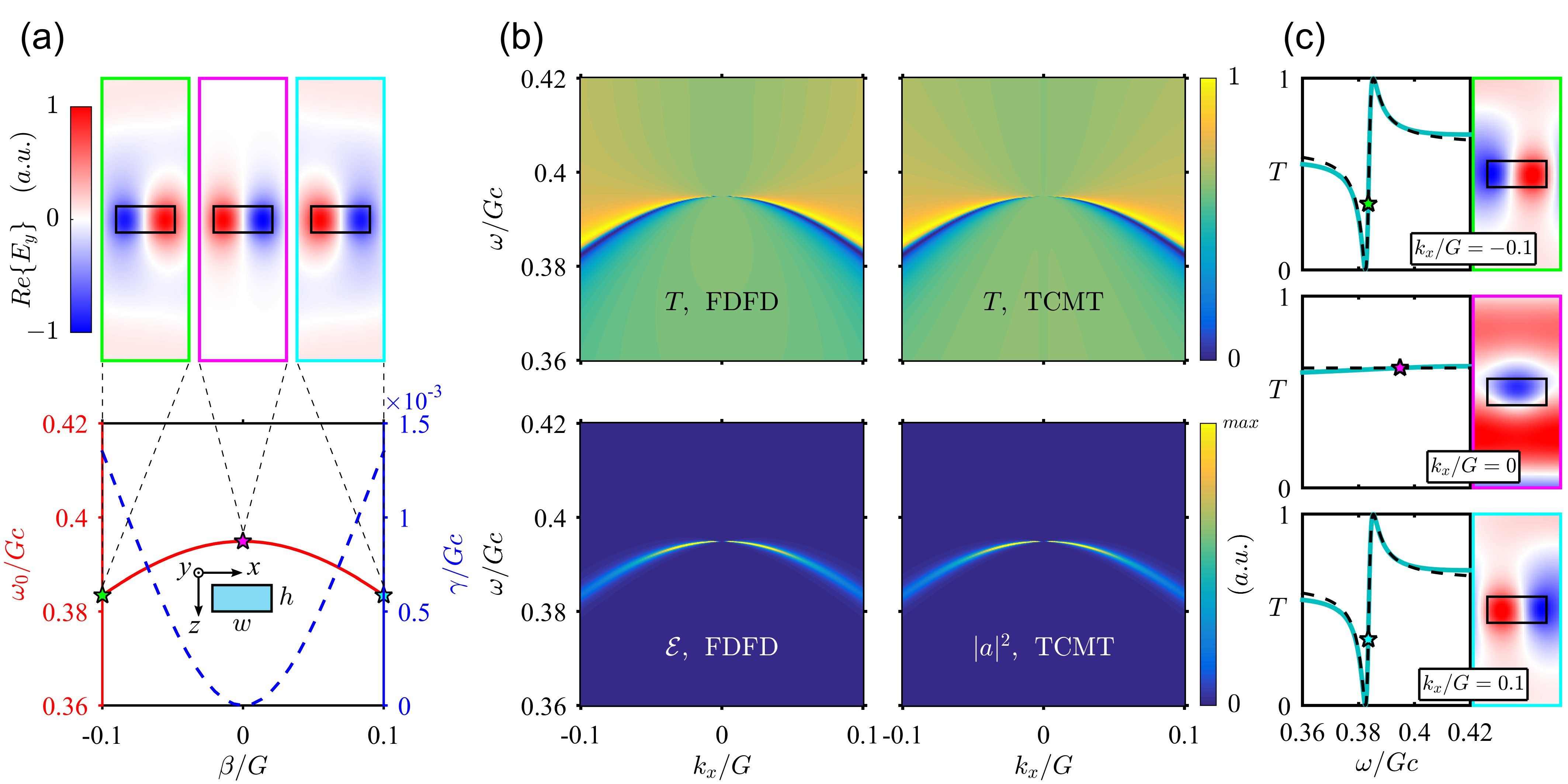}
     \caption{Transmission through the grating with rectangular unit cell (D$_{2h}$). (a) The spectra of the high-$Q$ leaky band calculated by FDFD with the mode profiles shown on top. (b) Transmittance and the energy stored by the FDFD and the TCMT methods. (c) Transmittance against the incident frequency; TCMT -- dotted line, FDFD -- solid line. The scattering solutions at $\omega=\omega_0$ are shown on the right of each plot. The geometry is described in the inset in subplot~(a); $w = 0.675 \Lambda$, $h = 0.3 \Lambda$.}
    \label{Fig4}
\end{figure*}
\section{Symmetries and numerical validation}\label{Sec5}
In this section, we present numerical results obtained by the FDFD in support of our findings~\cite{rumpf2022electromagnetic}. We will analyze metasurfaces (gratings) with various symmetries of unit cell according to the scheme shown in Fig.~\ref{Fig3}.  The used group-theoretical notations are taken from \cite{dresselhaus2010group}. 

In all simulations, a plane wave is incident from the half-space $z<0$. In accordance with Fig.~\ref{Fig2}, this means that the incident channels are $s_1^{\smallp}$ if $k_x>0$, and $s_3^{\smallp}$ if $k_x<0$.
For convenience in notation, we shall use the reciprocal lattice constant $G$, which is related to the lattice period $\Lambda$
\begin{equation}
G=\frac{2\pi}{\Lambda}.
\end{equation}

%The symmetry operations can be described with the use of the first Pauli matrix 
%\begin{equation}
%\hat{\sigma}_x=
%\left(
%\begin{array}{cc}
%0& 1 \\
%1& 0
%\end{array}\right)
%\end{equation}
%acting on the coupling vectors. Care is needed, though, as the channel functions are in general not mapped onto each other after the unitary transformation Eq.~\eqref{Udiag} that brings  $\widehat{C}_{\scs{4\!\times\!4}}$ to the simplified form given by Eq.~\eqref{C44}. This means that the non-zero elements of all matrix representations of the symmetry operation will acquire phase factors in accordance with Eq.~\eqref{Udiag}. Therefore, the conditions imposed by the symmetries can only be expressed by equating the  absolute values of the elements of the coupling vectors.

%%%%%%%%%%%%%--Symmetric---%%%%%%%%%%%%%%%%%%%%%%

%\newpage
%\
\subsection{Rectangular unit cell: D$_{2h}$ symmetry}
We start from the case of the grating with a rectangular unit cell. The spectrum of the real and imaginary parts of the leaky band resonant eigenfrequencies calculated by the FDFD method is shown in Fig.~\ref{Fig4}~(a). The profiles of the eigenmodes are shown on top of the subplot. Notice that in the $\Gamma$-point $\beta/G=0$ the imaginary part of the complex eigenfrequency equals zero. This indicates the presence of a symmetry protected BIC. The BICs of this type are omnipresent in dielectric gratings \cite{Bulgakov18} due to the Friedrich-Wintgen coupling \cite{Friedrich85} between two degenerate guided modes folded into the first Brillouin zone \cite{Hu2022}. Since the system retains two mirror symmetries, the coupling vector must simultaneously satisfy 
%\begin{align} 
%& {\bf \kappa}_{\smr,
%(\sml)}=\pm\hat{\sigma}_x{\bf \kappa}_{\smr,%(\sml)}, \nonumber \\
%& {\bf \kappa}_{\smr,(\sml)}={\bf \kappa}_{\sml,%(\smr)}.
%\end{align}
\begin{align} 
& |\varkappa_1|=|\varkappa_2|, \ |\varkappa_3|=|\varkappa_4|,  \nonumber \\
& |\varkappa_1|=|\varkappa_3|, \ |\varkappa_2|=|\varkappa_4|.
\end{align}
After applying the above equations to Eq.~\eqref{kappa_final} one finds that 
\begin{equation}
\alpha=\pm \pi/4, \ \delta=0.
\end{equation}
Since the eigenmode is symmetric with respect to inversion of the $z$-axis, we choose $\alpha=\pi/4$.
Thus, we are left with a single unknown parameter $\tau$ in applying the TCMT solutions Eqs.~(\ref{TT}--\ref{a23}). In general, the parameter $\tau$ can be dependent on the angle of incidence. In order to fit the TCMT formulas to the numerical spectra we interpolate $\alpha$ and $\tau$ in the range $k_x/G\in [0, 0.1]$ by the following formulas
\begin{align}\label{interp4}
&\alpha=\frac{\pi}{4}, \nonumber \\
&\tau = -0.790 - 0.113 \frac{k_x}{G}.
\end{align}
\begin{figure*}[t]
    \centering
    \includegraphics[width=1\linewidth]{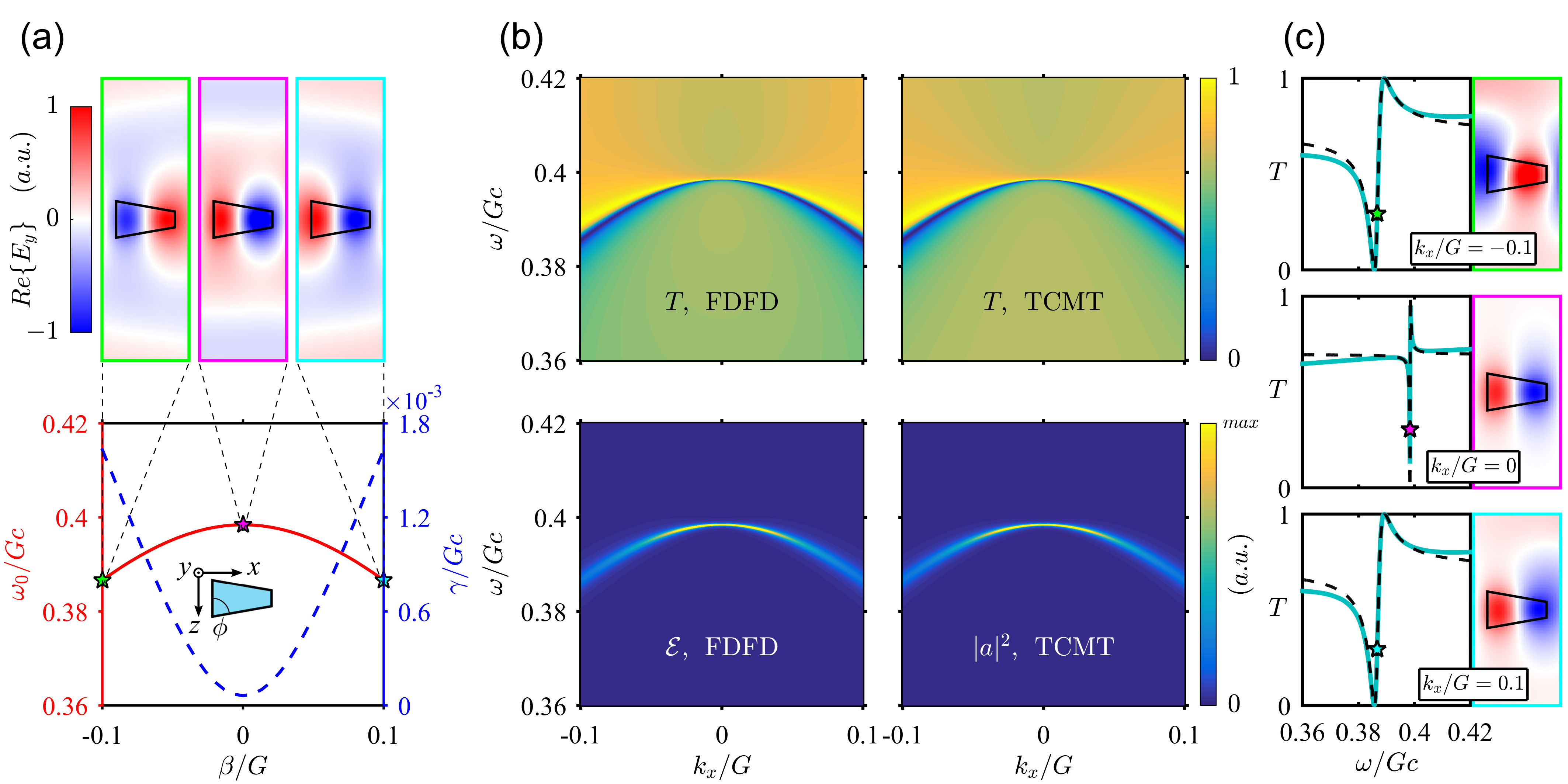}
    \caption{Transmission through the grating with isosceles trapezoid unit cell (C$_{2v}^x$). (a) The spectra of the high-$Q$ leaky band calculated by FDFD with the mode
profiles shown on top. (b) Transmittance and the energy stored by the FDFD and the TCMT methods. (c) Transmittance against the incident
frequency; TCMT – dotted line, FDFD – solid line. The geometry is described in the inset in subplot (a). The scattering solutions at $\omega=\omega_0$ are shown on the right of each plot. The in-plane mirror symmetry is broken by setting  $\phi = 80^{\circ}$ according to the scheme shown in Fig.~\ref{Fig3}.}
    \label{Fig5}
\end{figure*}
In the upper panel of Fig.~\ref{Fig4}~(b) we show the transmittance $T$ calculated by both the FDFD and the TCMT approaches. Note that now the spectra are plotted against the $x$-component of the wave vector $k_x$ in the surrounding medium. In full consistence with Fig.~\ref{Fig4}~(a) we observe a narrow Fano feature which collapses at the $\Gamma$-point. In the bottom panel of Fig.~\ref{Fig4}~(b) we compare the energy stored in the dielectric grating $\mathcal{E}$ calculated by FDFD against the squared amplitude of the resonant eigenmodes $|a|^2=|a_{\sml,\smr}|^2$ calculated by Eq.~\eqref{a14} and Eq.~\eqref{a23}.
In Fig.~\ref{Fig4}~(c) we demonstrate the profile of the Fano resonance at three different values of $k_x$ together with the scattering solutions at $\omega=\omega_0$. Notice that in the case of normal incidence $k_x/G=0$ the scattering field does not resemble the resonant mode profile from Fig.~\ref{Fig4}~(a). The reason for this is that the high-$Q$ leaky mode becomes a BIC in the $\Gamma$-point and the scattering solution is given by the background. The background is due to low-$Q$ frequency modes  \cite{Hu2022}, so one can clearly see that there is a residual dependence on $\omega$ that leads to a discrepancy between the TCMT and the FDFD methods. Also, notice that the transmittance reaches both zero and unity in the vicinity of the resonant eigenfrequency. In fact, this result is identical to the transmittance formula presented in \cite{Fan03}.
%%%%%%--Out-of-plane----%%%%%%%%
\subsection{Isosceles trapezoid  unit cell: C$_{2v}^{x}$ symmetry}
Now consider the situation when the in-plane mirror symmetry is broken whereas the out-of-plane mirror symmetry is retained. The spectrum of the real and imaginary parts of the leaky band resonant eigenfrequencies calculated by the FDFD method is shown in Fig~\ref{Fig5}~(a). The profiles of the eigenmodes are shown on top of the subplot. The symmetry is broken by changing the angle $\phi$ as shown in the inset in Fig.~\ref{Fig5}~(a).  Since the system does not have the $yz$-plane mirror symmetry the symmetry protected BIC is destroyed and the imaginary part of the resonant eigenfrequency does not equal zero in the $\Gamma$-point. The only condition that the coupling vector must satisfy is
\begin{align} 
 |\varkappa_1|=|\varkappa_2|, \ |\varkappa_3|=|\varkappa_4|.
\end{align}
Although we now have one less symmetry condition than in the case of rectangular unit cell (D$_{2h}$), Eq.~\eqref{kappa_final} results in the same solution 
\begin{equation}
\alpha=\pm\pi/4,\ \delta=0.
\end{equation}
As before, since the mode is symmetric we take $\alpha=\pi/4$.
The parameters are interpolated in the range $k_x/G\in [0, 0.1]$ by the following formula
\begin{align}\label{interp5}
& \alpha=\frac{\pi}{4}, \nonumber \\
& \tau = -0.834 - 0.113 \frac{k_x}{G}.
\end{align}
In the upper panel of Fig.~\ref{Fig5}~(b) we show the transmittance $T$ calculated by both the FDFD and the TCMT approaches. As before we observe a narrow Fano feature which, however, does not collapse in the $\Gamma$-point. The electromagnetic energy stored in the dielectric grating $\mathcal{E}$ calculated by the FDFD is compared with the squared amplitude of the resonant eigenmodes calculated by Eq.~\eqref{a14} and Eq.~\eqref{a23} in the lower panel of Fig.~\ref{Fig5}~(b).
\begin{figure*}[t]
\centering
\includegraphics[width=1\linewidth]{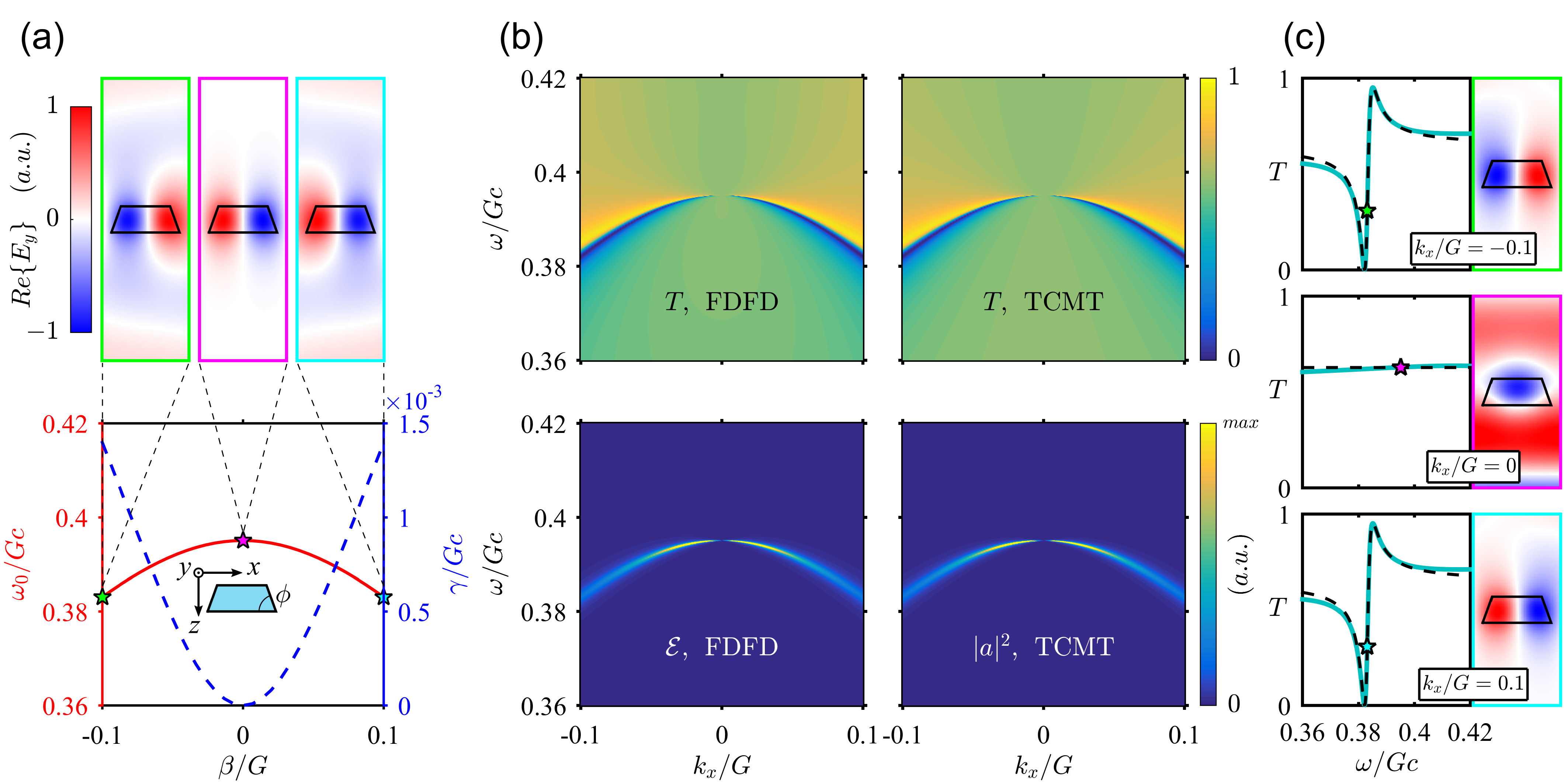}
\caption{Transmission through the grating with isosceles trapezoid unit cell (C$_{2v}^z$). (a) The spectra of the high-$Q$ leaky band calculated by FDFD with the mode
profiles shown on top. (b) Transmittance and the energy stored by the FDFD and the TCMT methods. (c) Transmittance against the incident
frequency; TCMT – dotted line, FDFD – solid line. The geometry is described in the inset in subplot (a). The scattering solutions at $\omega=\omega_0$ are shown on the right of each plot. The $yz$-plane mirror symmetry is broken by setting  $\phi = 70^{\circ}$ according to the scheme shown in Fig.~\ref{Fig3}.}
    \label{Fig6}
\end{figure*}
In Fig.~\ref{Fig5}~(c) we demonstrate the profile of the Fano resonance at three different values of $k_x$ together with the scattering solutions at $\omega=\omega_0$. Now in the case of normal incidence $k_x/G=0$ the scattering field looks similar to the resonant mode profile from Fig.~\ref{Fig5}~(a), since the high-$Q$ mode couples with incident light at all incident angles. In all other aspects the results are identical to those obtained for the structure with a rectangular unit cell (D$_{2h}$). Thus, we conclude that the only effect of breaking the $yz$-plane mirror symmetry while the $xy$-plane mirror symmetry is preserved is destruction of the symmetry protected BIC. We mention in passing that the interpolation coefficients in Eq.~\eqref{interp5} do not change significantly in comparison to the D$_{2h}$-symmetry case, Eq.~\eqref{interp4}. The direct process transmission amplitude $\tau$ is almost independent of the angle of incidence. 

%%%%%%%%%--In-plane----%%%%%%
%\pagebreak
\subsection{Isosceles trapezoid unit cell: C$_{2v}^{z}$ symmetry}
The $yz$-plane mirror symmetry requires that
\begin{equation}
|\varkappa_1|=|\varkappa_3|, \ |\varkappa_2|=|\varkappa_4|,
\end{equation}
{which together with Eq.~\eqref{kappa_final} leads to 
\begin{equation}
\delta=0.
\end{equation}}
Notice that now we have two independent parameters $\tau$ and $\alpha$ which have to be interpolated in the range $k_x/G \in [0, 0.1]$. 
The results of our numerical simulations are shown in Fig.~\ref{Fig6}. As before in subplot~(a) we demonstrate the dispersion of the high-$Q$ eigenmode with the eigenmode profiles shown above the dispersion curve. The two unknown parameters are interpolated by the formulas below 
\begin{align}\label{interp6}
& \tau = -0.792 - 0.113\frac{k_x}{G}, \notag \\
& \alpha = 1.24\frac{\pi}{4}.
\end{align}
The comparison between the TCMT and FDFD solutions in the whole interpolation range is shown in Fig.~\ref{Fig6}~(b). In Fig.~\ref{Fig6}~(c) are shown the profiles of Fano resonances with the corresponding scattering solutions at $\omega=\omega_0$ presented on the right of each plot. One can see that since the $yz$-plane mirror symmetry is present, the symmetry protected BIC is preserved. On the other hand, one can see in Fig.~\ref{Fig6}~(c) that the resonance does not reach unity at its peak. The analysis of Eq.~\eqref{TT} shows that a unity transmission at the peak of the Fano resonance is only possible with $\alpha=\mp\pi/4$ which would indicate the (anti)symmetry of the coupling with the scattering channels in the upper and lower half-spaces. This asymmetry in the coupling is a direct consequence of broken out-of-plane mirror symmetry. This case is important from the application view-point since in realistic setups the $xy$-plane mirror symmetry is broken by a substrate. It is worth mentioning that $\alpha=\pm\pi/4$ is not strictly forbidden by broken the $xy$-plane mirror symmetry. Yet, a robust reflection zero is difficult to engineer, since it only occurs in two points out of $\alpha\in[-\pi/2 , \pi/2]$ while the symmetry breaking favors solutions with asymmetric coupling $\alpha\neq\pm \pi/4$. 

%%%%%%--Inversion---%%%%%%%%%%%%%
\begin{figure*}[ht]
    \centering
    \includegraphics[width=1\linewidth]{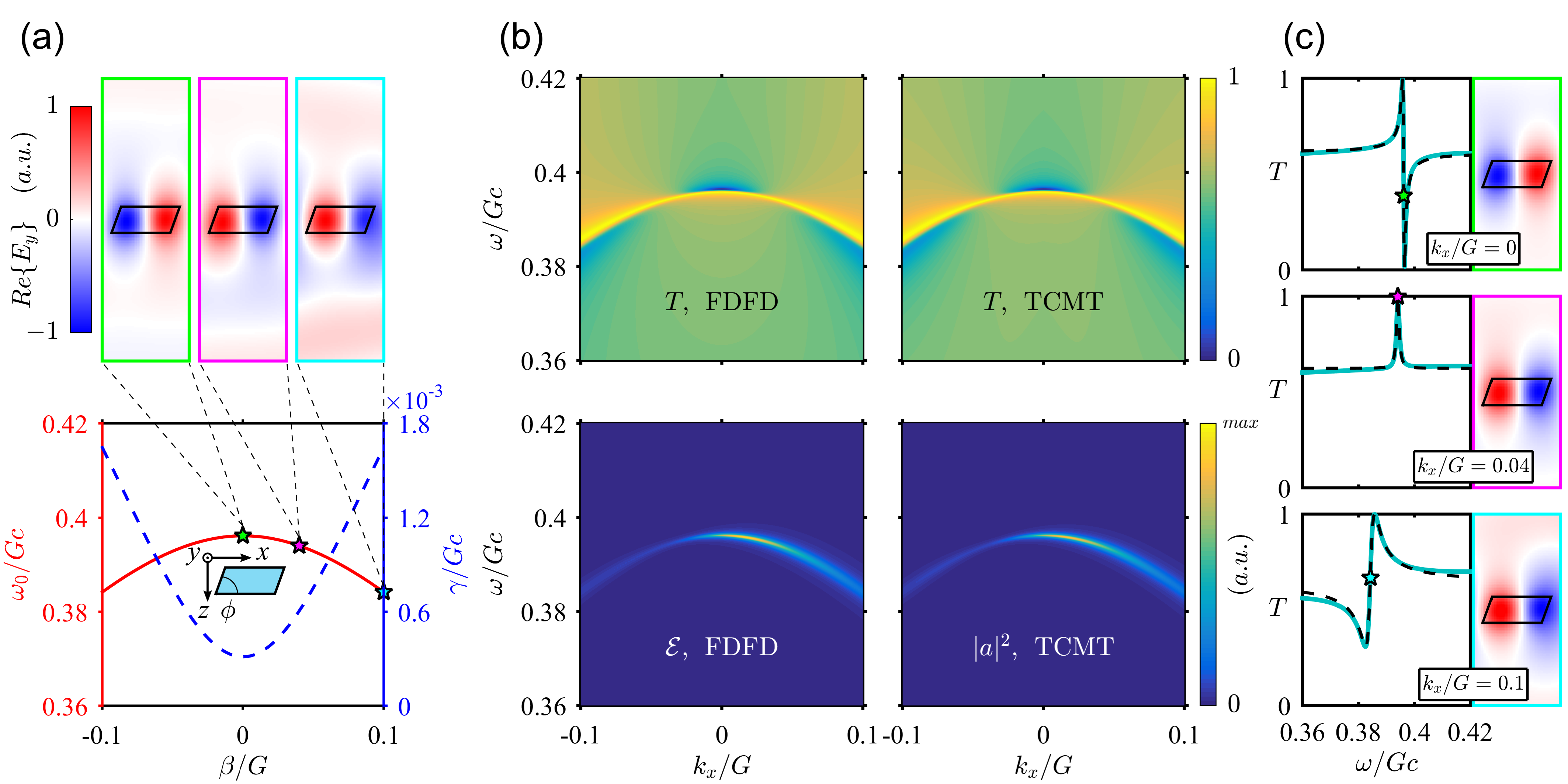}
    \caption{ Transmission thorough metasurface with parallelogram unit cell (C$_{2h}^y$). (a) The spectra of the high-$Q$ leaky band calculated by FDFD with the mode
profiles shown on top. (b) Transmittance and the energy stored by the FDFD and the TCMT methods. (c) Transmittance against the incident
frequency; TCMT – dotted line, FDFD – solid line. The geometry is described in the inset in subplot (a). The scattering solutions at $\omega=\omega_0$ are shown on the right of each plot. The $yz$-plane and $xy$-plane mirror symmetries are broken by setting  $\phi = 70^{\circ}$ according to the scheme shown in Fig.~\ref{Fig3}.}
    \label{Fig7}
\end{figure*}
\pagebreak
\subsection{Parallelogram unit cell: C$_{2h}^{y}$ symmetry}
This symmetry can be prescribed by simultaneously flipping the coupling vector and changing the propagation direction to yield 
\begin{equation}
|\varkappa_1|=|\varkappa_4|, \ |\varkappa_2|=|\varkappa_3| .
\end{equation}
The above equation together with Eq.~\eqref{kappa_final} results in
\begin{equation}
\alpha=\pm\frac{\pi}{4}.
\end{equation}
The spectrum of the radiative mode together with the field profiles is shown in Fig.~\ref{Fig7}~(a).
One can see in Fig.~\ref{Fig7}~(a) that because the $yz$-plane mirror symmetry is broken there is no symmetry protected BIC in the $\Gamma$-point.
Since the eigenmode is antisymmetric with respect to inversion, we take $\alpha=-\pi/4$. Thus, in the case of inversion symmetry we have two unknown parameters, $\tau$ and $\delta$, both dependent on the angle of incidence. In addition, at normal incidence we have ${\bf \kappa}_{\smr}={\bf \kappa}_{\sml}$ which dictates $\delta=0$ if $k_x/G=0$. Therefore, for applying the TCMT formulas, we interpolate the parameters as follows
\begin{align}
& \alpha=-\frac{\pi}{4}, \nonumber \\
& \tau = -0.782 - 0.170\frac{k_x}{G}, \nonumber \\
& \delta = 49.61\frac{k_x}{G} -258.5\left(\frac{k_x}{G}\right)^2.
\end{align}
In the upper panel Fig.~\ref{Fig7}~(b) we show the transmittance $T$ while the lower panel shows the energy $\cal{E}$ stored in the metasurface as functions of $k_x$ and $\omega$. Notice that, unlike the transmittance, the energy stored is asymmetric with respect to $k_x\rightarrow-k_x$. This is a consequence of $\delta\neq 0$ and $\bf{n}\neq\bf{n}^*$ in Eq.~\eqref{n_main}. Equivalently, one can state that ${\bf \kappa}_{\smr}\neq{\bf \kappa}_{\sml}$ if $k_x\neq 0$. According to Eq.~\eqref{coup_decoup} the coupling and decoupling vectors for the guided mode are not equal to each other, ${\bf \kappa}_{\sml,(\smr)}\neq {\bf d}_{\sml,(\smr)}$. The asymmetry between the coupling and decoupling coefficients manifests itself in the absence of transmittance zeros, see Fig.~\ref{Fig7}~(c), except at normal incidence where the coupling-decoupling symmetry is restored. It is also interesting to point out that, unlike to the previous cases where we found that both $\alpha$ and $\tau$ only slightly depend on the angle of incidence, in the case of the inversion symmetry the parameter $\delta$ varies significantly within the interpolation range. One can see in Fig.~\ref{Fig7}~(c) that the Fano resonance occurs in different shapes at different angles of incidence. Its profile can be flipped about the center-frequency, as it can be seen by comparing $k_x/G=0$ and $k_x/G=0.1$. The resonance is found to be symmetric (Lorentzian) at $k_x/G=0.04$ in which case $\delta=\pi/2$.

%%%%%%%-Asymetric---%%%%%%%%%%%%%%
\begin{figure*}[ht]
    \centering
    \includegraphics[width=1\linewidth]{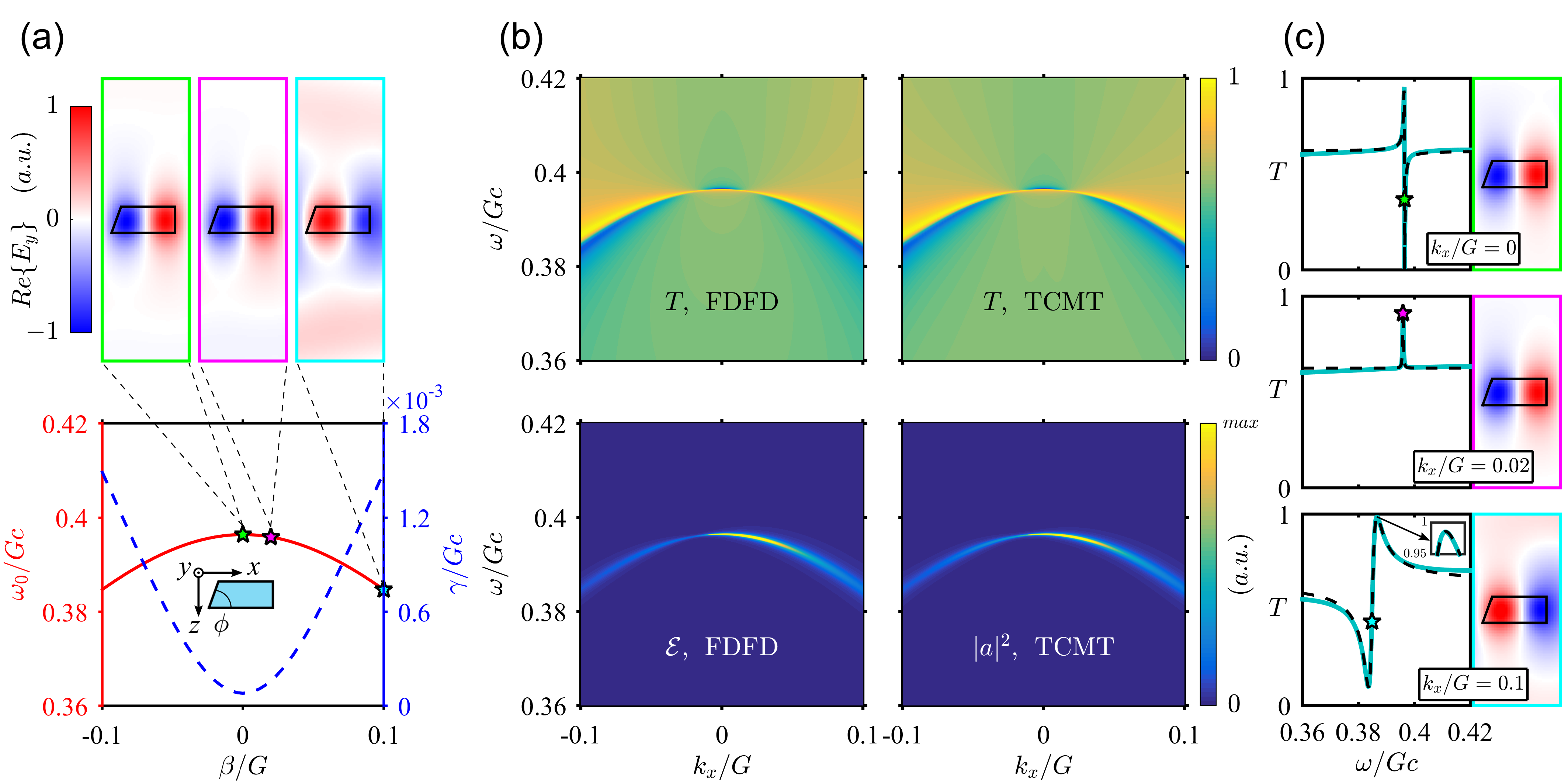}
    \caption{Transmission through the grating with right trapezoid unit cell (C$_s^y$). (a) The spectra of the high-$Q$ leaky band calculated by FDFD with the mode
profiles shown on top. (b) Transmittance and the energy stored by the FDFD and the TCMT methods. (c) Transmittance against the incident
frequency; TCMT – dotted line, FDFD – solid line. The geometry is described in the inset in subplot (a). The scattering solutions at $\omega=\omega_0$ are shown on the right of each plot. The $yz$-plane and $xy$-plane mirror symmetries are broken by setting  $\phi =70^{\circ}$ according to the scheme shown in Fig.~\ref{Fig3}.}
    \label{Fig8}
\end{figure*}
%\newpage
\subsection{Right trapezoid  unit cell: C$_{s}^{y}$ symmetry}
For a metasurface with a right trapezoid unit cell (C$_s^y$) there are no extra constraints on parameters in Eq.~\eqref{kappa_final} except ${\bf \kappa}_{\smr}={\bf \kappa}_{\sml}$ at normal incidence. Thus, we have to interpolate all three parameters $\tau, \ \alpha$, and $\delta$. The following formulas are used for interpolation in this subsection.
\begin{align}
& \tau = -0.787 - 0.119\frac{k_x}{G}, \nonumber \\ 
& \alpha =  -0.877 -0.032\frac{k_x}{G}, \nonumber \\
& \delta =  104.2\frac{k_x}{G} -1414\left(\frac{k_x}{G}\right)^2 
+6459 \left(\frac{k_x}{G}\right)^3. 
\end{align}
Notice that, as always, $\delta=0$ at normal incidence.  The spectrum of the real and the imaginary parts of the resonant eigenfrequency is shown in Fig.~\ref{Fig8}~(a). There is no in-$\Gamma$ symmetry protected BIC since the $yz$-plane mirror symmetry is broken. The simulation results for the transmittance and the energy stored are shown in Fig.~\ref{Fig8}~(b). Similarly to the case of the inversion symmetry the energy stored in the metasurface is asymmetric with respect to $k_x\rightarrow -k_x$ because in general $\delta\neq 0$. The profiles of Fano resonances are shown in Fig.~\ref{Fig8}~(c) for three different values of $k_x$. One can see in Fig.~\ref{Fig8}~(c) that, in the case of the asymmetric metasurface neither reflectance nor transmittance zeros are generally present in the spectra. The transmittance zeros only occur at $\delta=0$, i.e. at the normal incidence. The parameters $\alpha$ and $\delta$ vary considerably within the interpolation range to result in a variation of the shape of the Fano resonance. Accidental recovery of $\alpha=\pm\pi/4$, leading to reflectance zeros, as well as $\delta=0$, leading to transmittance zeros, is not forbidden by the absence of symmetry. Yet, transmittance and reflectance zeros could only be obtained by fine-tuning of parameters and, therefore, are not robust to fabrication inaccuracies. For example, in the case $k_z/G=0.1$ depicted in the lower panel of Fig.~\ref{Fig8}~(c) the transmittance almost reaches, but not exactly equal to, unity. As it has been mentioned in the introduction, the asymmetric metasurfaces have recently attracted attention as a platform for engineering UGRs. It is evident that the key to UGRs is asymmetry of decoupling vector which can be obtained at $\delta\neq 0$. In the next section we apply the TCMT approach to UGRs.

%%%%%%%%%%%%%%%%%%%%%%%%%%%%%%%%%%%%%%%%%%%%%%%%%%%%%%%%
%%%%%%%%%%%%%%%%%%%%%%%%%%%%%%%%%%%%%%%%%%%%%%%%%%%%%%%%%%
%%%%%%%%%%%%%%%%%%%%%%%%%%%%%%%%%%%%%%%%%%%%%%%%%%%%
\section{Uniguided and unicoupled resonant modes}\label{Sec6}
By definition, an UGR is decoupled from outgoing radiation in one of the two outgoing channels. Without a loss of generality we consider the situation when there is no outgoing radiation in channel $s_3^{\smallm}$, see Fig.~\ref{Fig2}. Then,  according to the definitions introduced in Eq.~\eqref{Dmat} for the right-going waves we should have $d_3=0$. By using Eq.~\eqref{coup_decoup} and Eq.~\eqref{kappa_final} one finds that $d_3=0$ and $\gamma\neq0$ result in two possible sets of conditions
\begin{align}\label{URGcond0}
& \tau=0, \ \ \alpha=0,
\end{align}
and
\begin{align}\label{URGcond}
& \delta=\frac{\pi}{2}, \nonumber \\
& \tan\alpha=\frac{\tau}{1+\rho}.
\end{align}
The first set of conditions, Eq.~\eqref{URGcond0}, can be satisfied only in metasurfaces with isosceles trapezoid (C$_{2v}^{z}$) or right trapezoid (C$_{s}^{y}$) shapes of the unit cell. The second set of conditions, Eq.~\eqref{URGcond}, can be satisfied in metasurfaces and metasurfaces with parallelogram (C$_{y}^{2h}$)  and right trapezoid (C$_{s}^{y}$) shapes of the unit cell. Note that in metasurfaces with $yz$-plane mirror symmetry a UGR can only occur on the zero-transmittance background, $\tau=0$, which is consistent with \cite{Zhang2022}. At the same time, since for metasurfaces with inversion symmetry $\alpha=\pm\pi/4$ the second line in Eq.~\eqref{URGcond} dictates that a UGR is only possible with zero-reflectance background, $\rho=0$, which is again consistent with \cite{Zhang2022}. Either of the two sets of conditions, Eq.~\eqref{URGcond0} and Eq.~\eqref{URGcond}, can be met by varying two geometric parameters. It is obviously a time-consuming computational task since two equations have to be satisfied simultaneously or, in the other words, UGRs are non-robust to variation of geometry.

\begin{figure}[th]
    \centering
    \includegraphics[width=0.40\textwidth]{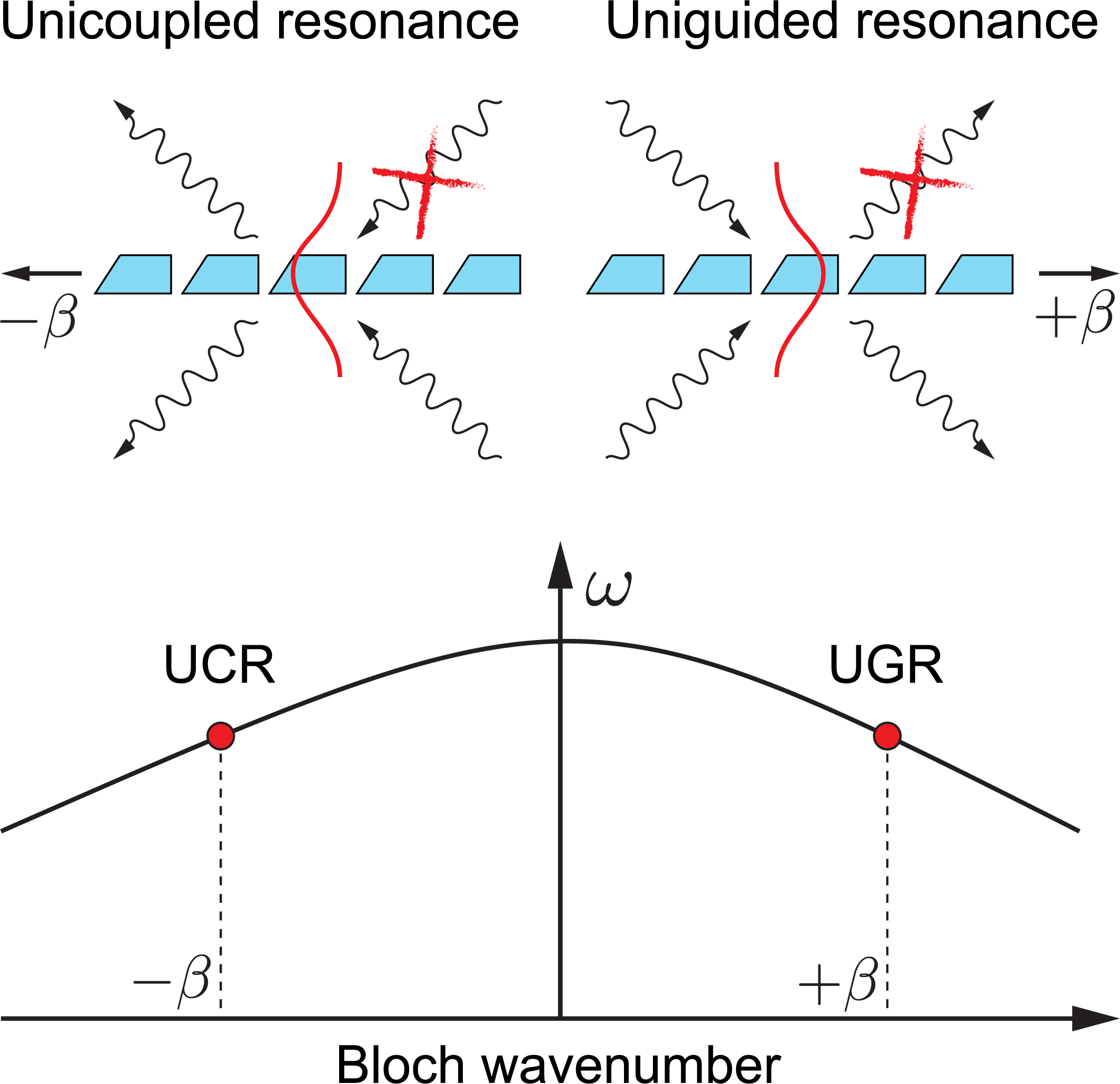}
    \caption{Duality between the unidirectional coupled resonant mode (UCR) and unidirectional guided resonance (UGR) in a grating with C$^y_s$-symmetry of the unit cell.}
    \label{Fig15}
\end{figure}

In a metasruface with $yz$-plane mirror symmetry ${\bf d}_{\smr}={\bf\kappa}_{\smr}$, so $\varkappa_1=0$, which means that the UGR is not only decoupled from outgoing light in a certain channel, but also can not be excited by incident light on the same side. The same conclusion holds true for the mode counter-propagating to the UGR. Thus, UGRs occur in pairs $d_3=0,\ \varkappa_1=0$ and $d_1=0,\ \varkappa_3=0$.
In the case of the inversion symmetry the situation is similar, but, according to Eq.~\eqref{kappa_final}, the zero coupling occurs on the other side of the metasurface $\varkappa_2=0$. The UGRs again emerge in pairs but now the zero-coupling and the zero-decoupling directions are swapped between the counter-propagating modes, i.e. we have $d_3=0, \ \varkappa_2=0$ and $d_2=0, \ \varkappa_3=0$. The picture becomes more interesting when all symmetries are lifted and Eq.~\eqref{URGcond} is satisfied. One can easily see that if $d_3=0$ and $\alpha\neq\pm\pi/4$, then $\varkappa_{1,2}\neq 0$. Thus, the UGR is coupled with incident waves on both sides of the metasurface. We, however, remember that the coupling and decoupling vectors are  swapped when the propagating direction is reversed. Therefore, we always have $\varkappa_3=d_3$. Thus, in an asymmetric metasurface an UGR is dual to a counter-propagating mode decoupled from incident radiation from one of the sides of the metasurfaces. We suggest to term this type of mode {\it unidirectional coupled resonant mode} (UCR). Notice, that, unlike UGR, the UCR can radiate into both half-spaces. This duality between UCR and UGR in a grating with C$^y_s$-symmetry of the unit cell is illustrated in Fig.~\ref{Fig15}.  

\begin{figure}[th]
    \centering
    \includegraphics[width=0.40\textwidth]{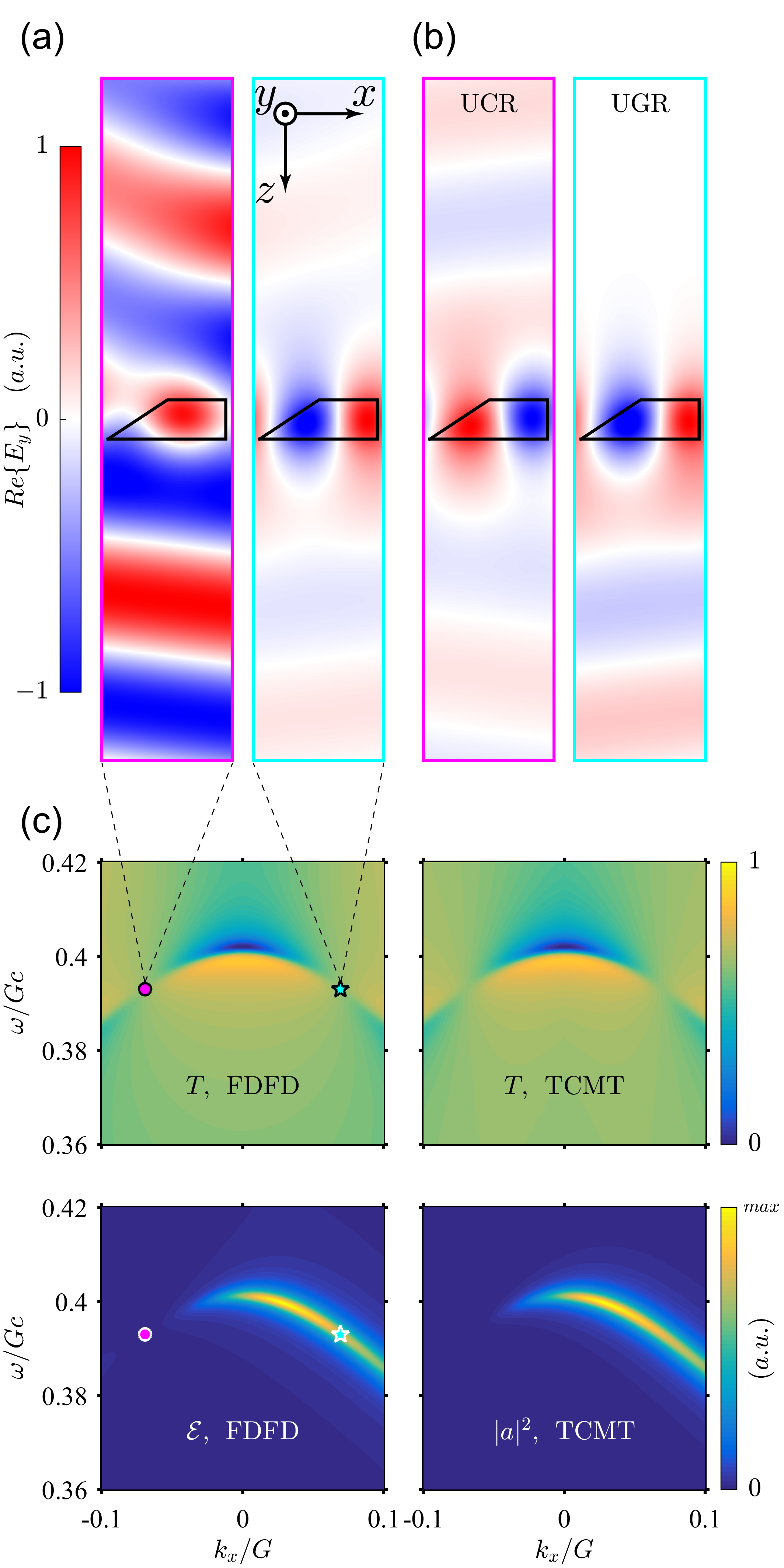}
    \caption{Uniguided and unicoupled resonances in metasurface with right trapezoid unit cell (C$_s^y$). The $yz$-plane and $xy$-plane mirror symmetries are broken by setting  $\phi = 33.42^{\circ}$ according to Fig.~\ref{Fig3}. The spectral position of UGR (UCR) is given by $\omega_0/Gc = 0.3939, \ k_x/G =\pm 0.0708$. (a) The scattering solution in the points of the UGR and the UCR. (b) The field profiles of the UGR and the UCR eigenmodes. (c) Comparison between the FDFD simulation and the TCMT models for transmittance $T$ and the energy stored $\cal{E}$. The position of the UGR is shown by a green star. The position of the UCR is shown by a red dot. }
    \label{Fig9}
\end{figure}

To verify our finding we applied the TCMT model to a UGR in the asymmetric metasurface from the previous section. First, an UGR was found by varying the geometry of the elementary cell in the point $\omega_0/Gc = 0.3939, \ k_x/G = 0.0708$. Second, the numerical spectra were compared with the TCMT predictions. The results are shown in Fig.~\ref{Fig9}. In Fig.~\ref{Fig9}~(a) we show the scattering solutions in the points of the UGR and UCR. In Fig.~\ref{Fig9}~(b) we demonstrated the UGR and UCR mode shapes. Finally, in Fig.~\ref{Fig9}~(c) we compare the numerical results for the transmittance and the energy stored with the TCMT model. One can see in Fig.~\ref{Fig9}~(b) that the UGR is a right-going wave decoupled from the channel $s_3^{\scs{\smallm}}$. The corresponding scattering solution is found by illuminating the system through channel $s_1^{\smallp}$. Note that in the case of UGR the scattering solution is similar to the UGR eigenmode and a resonant field enhancement is visible in the spectrum of the energy stored. One the other hand, the UCR, i.e. the left-going mode counter-propagating to the UGR, radiates to both sides of the metasurface, see Fig.~\ref{Fig9}~(b). The UCR is, however, not excited by a plane wave incident in channel $s_3^{\smallp}$ indicating a unidirectional coupling to incident light, $\varkappa_3=0$.
One can also see in Fig.~\ref{Fig9}~(c) that the TCMT model reproduces the spectra to a good accuracy. The parameters were interpolated by the formulas
\begin{align}
& \tau =      -0.776 -0.299 \frac{k_x}{G}, \nonumber \\
& \alpha =  -0.487 + 0.345\frac{k_x}{G}, \nonumber \\
& \delta =  31.57\frac{k_x}{G} -128.8\left(\frac{k_x}{G}\right)^2, 
\end{align}
that satisfy Eq.~\eqref{URGcond} in the UGR point.

For the sake of completeness, it is worth writing down the exact analytic expression for the $S$-matrix at the UGR point according to Eq.~\eqref{right-going}. In doing so, one should remember that Eq.~\eqref{right-going} is obtained after the unitary transformation Eq.~\eqref{Udiag} that redefined the scattering channels. After returning to the most generic expression for the $S$-matrix, Eq.~\eqref{0A3} and Eq.~\eqref{S2gen}, one can write all non-zero elements of the $S$-matrix as follows
\begin{align}\label{SUGR}
& S_{41}=S_{14}=i\tau e^{i(\psi+\xi)}\left[\frac{i(\omega_0-\omega)-\gamma}{i(\omega_0-\omega)+\gamma} \right], \nonumber \\
& S_{42}=S_{24}=\rho e^{i(\psi-\eta)}\left[\frac{i(\omega_0-\omega)-\gamma}{i(\omega_0-\omega)+\gamma} \right], \nonumber \\
& S_{31}=S_{13}=\rho e^{i(\psi+\eta)}, \nonumber \\
& S_{32}=S_{23}=i\tau e^{i(\psi-\xi)}.
\end{align}
Note that the absolute values of all matrix elements are independent of frequency. Thus, only the phase of the scattered wave is frequency dependent if a UGR or a UCR is excited. This phenomenon has been applied for engineering 
intensity flattened phase shifting in \cite{Zhang2021, Zhang2022, Li2023}.
%%%%%%%%%%%%%%%%%%%%%%%%%%%%%%%%%%%%%%%%
%%%%%%%%%--Summary----%%%%%%%%%%%%%%%%%%
%%%%%%%%%%%%%%%%%%%%%%%%%%%%%%%%%%%%%

%\widetext{
\begin{table*}[t]
\begin{tabular}{ |l|c|c|c|c|c|}
\hline
\multirow{3}{*}%{Symmetry} 
&  & & & & \\ 
Symmetry &  \ \ \ \ Rectangular \ \ \ \ 	&Isosceles   & \ \ \ Isosceles \ \ \  & \ \ \  \ \ \ \ Parallelogram  \ \ \ \ \ \ \ & \ \  Right    \ \ \\
                 &  \ \ \ \ \ (D$_{2h}$) \ \ \ \ 			&trapezoid (C$_{2v}^x$)   & \ \ \ trapezoid (C$_{2v}^z$) \ \ \  & \ \ \  \ \ \ \ (C$_{2h}^y$)  \ \ \ \ \ \ \ & \ \  trapezoid (C$_{s}^y$)    \ \ \\
& \ \ \ \  \includegraphics[width=2cm]{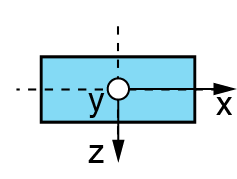}  \ \ \ \ &\ \ \  \includegraphics[width=2cm]{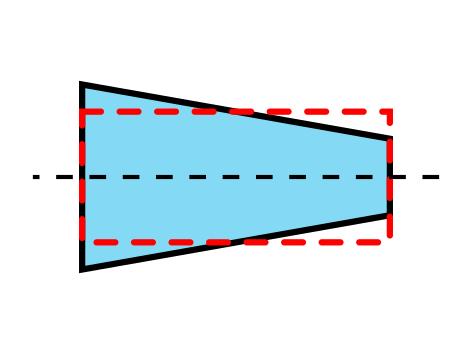}  \ \ \ & \includegraphics[width=2cm]{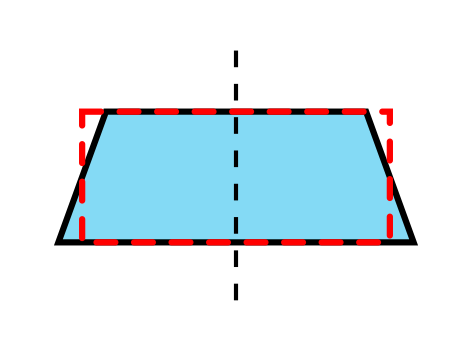}  & 
 \ \ \  \ \ \ \ \includegraphics[width=2cm]{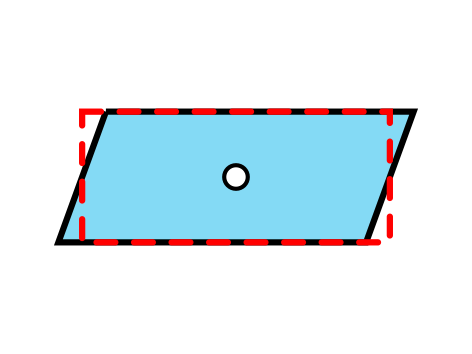}  \ \ \ \ \ \ \ & \ \   \includegraphics[width=2cm]{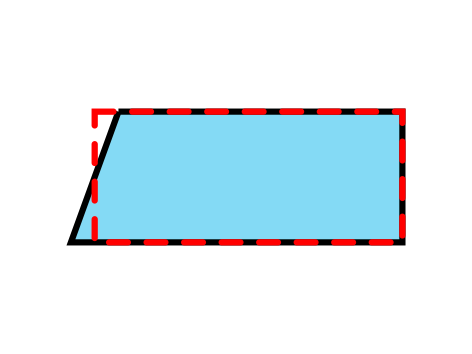}   \ \ 
 \\ 

\hline
%\multirow{3}{*}{Coupling constraint} & & & & %\\ & ${\bf \kappa}_{\sml,(\smr)}=\pm{\bf %\kappa}_{\smr,(\sml)}$ & ${\bf %\kappa}_{\sml,%(\smr)}=\pm\hat{\sigma}_x{\bf %\kappa}_{\sml,(\smr)}$ & ${\bf %\kappa}_{\sml,%(\smr)}=\pm\hat{\sigma}_x{\bf %\kappa}_{\smr,(\sml)}$ & N/A\\ 
%& & & &  \\
%\hline
\multirow{3}{*}
& & & & & \\
& $\tau\in[-1, 1],$ & $\tau\in[-1, 1]$, & $\tau\in[-1, 1]$, & $\tau\in[-1, 1]$, & $\tau\in[-1, 1]$, \\
{Parameters in Eq.~\eqref{TT},} & $\delta=0$, & $\delta=0$, & $\delta=0$,& $\delta\in[0, \pi)$, & $\delta\in[0, \pi)$,  \\
oblique incidence & $\alpha=\pm\pi/4$ &  $\alpha=\pm{\pi}/{4}$ & $\alpha\in[-\pi/2, \pi/2]$ & $\alpha=\pm{\pi}/{4}$ &  \ $\alpha\in[-\pi/2, \pi/2]$ \\
& & & & & \\
\hline
\multirow{3}{*}
& & & & & \\
& $\tau\in[-1, 1],$ & $\tau\in[-1, 1]$, & $\tau\in[-1, 1]$, & $\tau\in[-1, 1]$, & $\tau\in[-1, 1]$, \\
{Parameters in Eq.~\eqref{TT},} & $\delta=0$, & $\delta=0$, & $\delta=0$,& $\delta=0$, & $\delta=0$,  \\
normal incidence & $\alpha=\pm\pi/4$ &  $\alpha=\pm{\pi}/{4}$ & $\alpha\in[-\pi/2, \pi/2]$ & $\alpha=\pm{\pi}/{4}$ &  \ $\alpha\in[-\pi/2, \pi/2]$ \\
& & & & & \\
\hline
\multirow{3}{*}{Symmetry protected BIC} & & & & &\\ & $+$ & $-$ & $+$ & $-$ &  $-$ \\
& & & & & \\
\hline
\multirow{3}{*}{Robust reflectance zeros} & & & & &\\ &$+$ & $+$ & $-$ & $+$ &  $-$ \\
& & & & & \\
\hline
\multirow{3}{*}{Robust transmittance zeros} & & & & &\\ 
& $+$& $+$ & $+$ 	& only at normal  	&  only at normal  \\
&\ & \ & \  			& incidence 		&   incidence \\

\hline
\multirow{3}{*}{Uniguided resonance, UGR} & & & & &\\ 
& $-$ & $-$ 	&  possible if 	& possible if 	&  possible,   \\
& \ & \  		&   $\tau=0$ 	&  $|\tau|=1$	&  see Eqs.~\eqref{URGcond0},~\eqref{URGcond}  \\
\hline
\multirow{3}{*}{Unicoupled resonance, UCR} & & & & &\\ 
& $-$ & $-$ 	& coincides	&coincides 	&  counter-propagating\\
& \  & \  		& with UGR	&with UGR	&  to UGR\\
\hline
\end{tabular}
\caption{Allowed ranges of the parameters $\tau$, $\delta$, and $\alpha$ in
  Eq.~\eqref{TT} for the main unit-cell symmetries of a subwavelength dielectric grating shown in Fig.~\ref{Fig3}. The last five rows indicate whether a given symmetry permits ('$+$') or forbids ('$-$') symmetry-protected BICs, robust reflection/transmission zeros of Fano resonance, uniguided resonances (UGR), and unicoupled resonances (UCR).}\label{tab}
\end{table*}
\section{Summary and conclusions}\label{Sec7}
We proposed a temporal coupled mode theory (TCMT) for resonant response from quasi-guided modes in periodic dielectric metasurfaces. A generic set of constraints imposed onto the parameters of the temporal coupled mode theory by energy conservation and time-reversal symmetry is derived. It is demonstrated that the proposed approach allows for asymmetry between the coupling and decoupling coefficients. In particular, the TCMT approach is applied to the problem of Fano resonances induced by isolated quasi-guided modes in the regime of specular reflection. We derived a generic formula, Eq.~\eqref{TT}, for the line-shape of the Fano resonance in transmittance for the lossless metasurfaces in the framework of 2D electrodynamics. {An analytic expression for the Fano asymmetry parameter, Eq.~\eqref{asymmetry}, is found. It is demonstrated that our approach is applicable to all possible symmetries of the metasurface elementary cell. In particular, the approach explains the effects of the symmetry on the profile of the Fano resonance induced by an isolated high-$Q$ mode. It is shown that the proposed approach correctly describes the presence of robust reflection and transmission zeros in the spectra as well as the spectral signatures of bound states in the continuum (BICs). 

Thus far, there have been many attempts to derive the line shapes of the Fano resonances relying on the spectra of radiative modes \cite{Blanchard16, Alpeggiani17, Popov1986, Fehrembach2002, Gippius2005, Wang2013, Bykov15, Wu2022, Yuan2022} including systems with reflectionless scattering modes \cite{Zhou2016, Sweeney2020}.  
In particular, the TCMT was applied for analyzing the fundamental bounds on decay rates in asymmetric single-mode optical resonators with two scattering channels in \cite{Wang2013}, where it was shown that full reflection is always achievable at a Fano resonance, even for structures lacking mirror symmetries, whereas full transmission can only be seen in a symmetric configuration where the two decay rates are equal. In the context of the present work, the above result corresponds to normal incidence at which transmittance is always zero at the minimum of the resonant line if the high-$Q$ mode is excited. At the same time reflectance zeros are only possible if either inversion or $yz$-plane mirror symmetries are preserved. Our approach also complies with the earlier work on photonic crystal slabs \cite{Gippius2005} where the authors reported that reflectivity peaks to unity near the quasiguided mode resonance for normal light incidence in the absence of diffraction, depolarisation, and
absorptive losses, while for the oblique incidence the full reflectivity is reached only in symmetrical systems.  Interestingly, only $yz$-plane mirror symmetry does not guaranty the absence of reflectance zeros which can be obtained by fine-tuning the system's parameters \cite{Yuan2022}, such reflectance zeros are, however, non-robust.

For oblique incidence it was demonstrated many times \cite{Popov1986, Shipman2012, Bykov15} that both reflectance and transmittance zeros simultaneously occur in symmetric metasurfaces. More recently, Zhou and co-workers proposed a TCMT model that can take into account all possible symmetries of the elementary cell~\cite{Zhou2016}. Similarly to the model presented, their approach also implies asymmetry between the coupling and the decoupling matrices that manifests itself in the relationship $\widehat{D}=\widehat{K}\hat{\sigma}_x$, where where $\sigma_x$ is the Pauli-x matrix.  This equality follows from the time-reversal operation, which swaps the counter-propagating resonances with Bloch wavenumbers  $\beta$ and $-\beta$. Moreover, the authors came to the same conclusion that the inversion symmetry only allows for robust reflectance zeros at oblique incidence, meanwhile the $yz$-plane mirror symmetry only permits robust zeros in transmittance.
Nethertheless, the TCMT solution for reflectance derived in~\cite{Zhou2016}, is not generic as it does not reproduce the frequency independent transmittance/reflectance in the point unidirectional guided resonance (UGR) \cite{comment11}. The approach presented here is
found to be consistent with the existence of UGRs. It is shown to correctly describe the transmittance and resonant field enhancement in the spectral vicinity of an UGR in metasurfaces with an asymmetric elementary cell. The analytic expression for the $S$-matrix in the UGR point is found to be consistent with the previously reported effect of intensity flattened phase shifting \cite{Zhang2021, Zhang2022, Li2023}.
Furthermore, in the case of asymmetric metasurfaces the theory predicts that a UGR is dual to a counter-propagating mode of a peculiar type which is coupled with the outgoing wave on both sides of the metasurface but, nonetheless, exhibits only a single-sided coupling with incident waves. This type of mode, which we termed a unidirectional coupled resonance (UCR) is confirmed numerically. 

The discussion is summarized in TABLE I. We believe that the results presented can be of use in engineering Fano response from dielectric metasurfaces as well as pave a way for application of TCMT in nanophotonics. In particular, generalizations for lossy systems are required for describing enhanced light-matter interaction in the critical coupling regime. 

\begin{acknowledgements}
This work received financial support from the Ministry of Science and Higher Education of Russian Federation
(Project No. FSRZ-2023-0006). DNM appreciates discussions with Alexander E. Ershov.
\end{acknowledgements}
%%%%%%%%%%--Bibliography---%%%%%%%%%%%%%%%%%%%%%%

\bibliography{bib-PRX}% Produces the bibliography via BibTeX.

\appendix

\end{document}